\documentclass[12pt,habil,bindborder,english,secnumarabic,eqsecnum,nofootinbib]{tuthesis}
\usepackage{graphicx}
\usepackage{amssymb,amsfonts,amsmath,bm,theorem}
\usepackage[dvips,colorlinks=true,linkcolor=blue]{hyperref}
\usepackage{mathrsfs}
\usepackage{wrapfig}
\newcommand{\pard}[2]{\frac{\partial #1}{\partial #2}}

\newcommand{\tl}[1]{\tilde{#1}}

\newcommand{\bc}{\begin{center}}
\newcommand{\ec}{\end{center}}
\newcommand{\be}{\begin{equation}}
\newcommand{\ee}{\end{equation}}
\newcommand{\ba}{\begin{eqnarray*}}
\newcommand{\ea}{\end{eqnarray*}}
\newcommand{\bna}{\begin{eqnarray}}
\newcommand{\ena}{\end{eqnarray}}

\newcommand{\mpaa}{\begin{minipage}[t]{7.5cm}}
\newcommand{\mpea}{\end{minipage}}
\newcommand{\half}{\frac{1}{2}}

\newcommand{\ud}{\mathrm{d}}
\newcommand{\nl}{\mbox{}\\}

\newtheorem {defi}{Definition}

\newtheorem{theo}{Theorem} 
 
{\theorembodyfont{\rmfamily}\theoremstyle{plain}
\newtheorem {remark}{Remark} 
 
\newtheorem{example}{Example}}
{\theorembodyfont{\rmfamily}\theoremstyle{break}
}
\newcommand{\qed}{\phantom{xxxxxx}\hfill q.e.d.}

\begin{document}

\bc
{\LARGE \bf From Deterministic Chaos\\[1ex] 
to Anomalous Diffusion}\\[2ex]
{\Large Dr.\ habil.\ Rainer Klages\\
Queen Mary University of London\\
School of Mathematical Sciences\\
Mile End Road, London E1 4NS\\
e-mail: r.klages@qmul.ac.uk}
\ec

  
\setcounter{tocdepth}{3}
\tableofcontents

\newpage
  

\chapter{Introduction}

Over the past few decades it was realized that deterministic dynamical
systems involving only a few variables can exhibit {\em complexity}
reminiscent of many-particle systems if the dynamics is {\em chaotic},
as can be quantified by the existence of a positive Ljapunov exponent
\cite{Schu}. Such systems provided important paradigms in order to
construct a theory of nonequilibrium statistical physics starting from
first principles, i.e., based on microscopic nonlinear equations of
motion. This novel approach led to the discovery of fundamental
relations characterizing transport in terms of deterministic chaos, of
which formulas relating deterministic diffusion to differences between
Ljapunov exponents and dynamical entropies form important examples
\cite{Do99,Gasp,Kla06}.  More recently scientists learned that
random-looking evolution in time and space also occurs under
conditions that are weaker than requiring a positive Ljapunov
exponent, which means that the separation of nearby trajectories is
weaker than exponential \cite{Kla06}. This class of dynamical systems
is called {\em weakly chaotic} and typically leads to transport
processes that require descriptions going beyond standard methods of
statistical mechanics. A paradigmatic example is the phenomenon of
anomalous diffusion, where the mean square displacement of an ensemble
of particles does not grow linearly in the long-time limit, as in case
of ordinary Brownian motion, but nonlinearly in time. Such {\em
anomalous tranport phenomena} do not only pose new fundamental
questions to theorists but were also observed in a large number
of experiments \cite{KRS08}.

This review gives a very tutorial introduction to all these topics in
form of three chapters: Chapter~\ref{prelim} reminds of two basic
concepts quantifying {\em deterministic chaos} in dynamical systems,
which are Ljapunov exponents and dynamical entropies. These approaches
will be illustrated by studying simple one-dimensional maps. Slight
generalizations of these maps will be used in
Chapter~\ref{chap:detdif} in order to motivate the problem of {\em
deterministic diffusion}. Their analysis will yield an exact formula
expressing diffusion in terms of deterministic chaos. In the first
part of Chapter~\ref{chap:anodif} we further generalize these simple
maps such that they exhibit {\em anomalous diffusion}. This dynamics
can be analyzed by applying continuous time random walk theory, a
stochastic approach that leads to generalizations of ordinary laws of
diffusion of which we derive a fractional diffusion equation as an
example. Finally, we demonstrate the relevance of these theoretical
concepts to experiments by studying the anomalous dynamics of
biological cells migration.

The degree of difficulty of the material presented increases from
chapter to chapter, and the style of our presentation changes
accordingly: While Chapter~\ref{prelim} mainly elaborates on textbook
material of chaotic dynamical systems \cite{Ott,Beck},
Chapter~\ref{chap:detdif} covers advanced topics that emerged in
research over the past twenty years \cite{Do99,Gasp,RKdiss}. Both
these chapters were successfully taught twice to first year Ph.D.\
students in form of five one-hour lectures. Chapter~\ref{chap:anodif}
covers topics that were presented by the author in two one-hour
seminar talks and are closely related to recently published research
articles \cite{KCKSG06,KKCSG07,DKPS08}.

\chapter{Deterministic chaos}\label{prelim}

To clarify the general setting, we start with a brief reminder about
the dynamics of time-discrete one-dimensional dynamical systems. We
then quantify chaos in terms of Ljapunov exponents and (metric)
entropies by focusing on systems that are closed on the unit
interval. These ideas are then generalized to the case of open
systems, where particles can escape (in terms of absorbing boundary
conditions). Most of the concepts we are going to introduce carry
over, suitably generalized, to higher-dimensional and time-continuous
dynamical systems.\footnote{The first two sections draw on
Ref.~\cite{dslnotes}, in case the reader needs a more detailed
introduction.}

\section{Dynamics of simple maps}  

\begin{figure}[htp]
    \begin{center}
    \includegraphics[width=15cm]{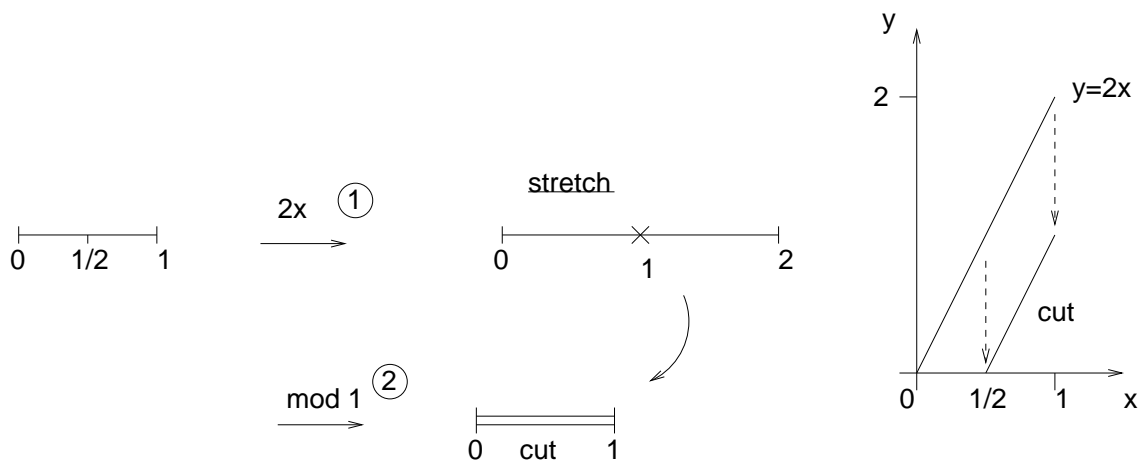}
    \caption{Stretch-and-cut mechanism in the Bernoulli shift.}
    \label{bernstretchfig}
    \end{center}
    \end{figure}

As a warm-up, let us recall the following:

\begin{defi} Let $J\subseteq \mathbb{R}, x_n \in
J, n\in\mathbb{Z}.$ Then 
\begin{equation}\label{mdef}
F:J\to J\quad ,\quad x_{n+1}=F(x_n)
\end{equation}
is called a {\em one-dimensional time-discrete map}.
$x_{n+1}=F(x_n)$ are sometimes called the {\em
equations of motion} of the dynamical system.  
\end{defi}
Choosing the initial condition $x_0$ {\em determines} the outcome
after $n$ discrete time steps, hence we speak of a {\em deterministic
dynamical system}. It works as follows:
\begin{eqnarray}
x_1&=&F(x_0) = F^1(x_0), \nonumber \\
x_2&=&F(x_1) = F(F(x_0)) =F^2(x_0) . \nonumber
\end{eqnarray}
\be
\Rightarrow\:F^m(x_0) :=\underbrace{F\circ F\circ\cdots F(x_0)}_{\mbox{m-fold composed map}} \label{eq:mfoldcomp}\quad.
\ee
In other words, there exists a unique solution to the equations of
motion in form of $x_n=F(x_{n-1})=\ldots=F^n(x_0)$, which is the
counterpart of the flow for time-continuous systems.  In the first two
chapters we will focus on simple piecewise linear maps. The following
one serves as a paradigmatic example \cite{Schu,Ott,ASY97,Do99}:

\begin{example} The Bernoulli shift (also shift map, doubling map,
dyadic transformation)

    \begin{figure}[htp]
    \begin{center}
    \includegraphics[width=5cm]{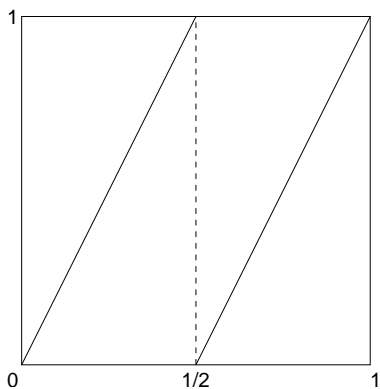}
    \caption{The Bernoulli shift.}
    \label{bernoullifig}
    \end{center}
    \end{figure}

The Bernoulli shift shown in Fig.~\ref{bernoullifig} can be defined by
\be
B:[0,1)\to[0,1)\;,\;B(x):=2x \;\mbox{mod 1}\; = \left\{\begin{array}{l} 2x\;,\; 0\le x < 1/2 \\ 
2x-1\;,\; 1/2 \le x < 1 \end{array}\right. \quad . \label{eq:bern}
\ee
\end{example}
One may think about the dynamics of such maps as follows, see
Fig.~\ref{bernstretchfig}: Assume we fill the whole unit interval with
a uniform distribution of points. We may now decompose the action of
the Bernoulli shift into two steps:

\begin{enumerate}
\item The map {\em stretches} the whole distribution of points by a
factor of two, which leads to {\em divergence} of nearby
trajectories. 
\item Then we {\em cut} the resulting line segment in the middle due
to the the modulo operation $\mod 1$, which leads to motion {\em
bounded} on the unit interval.  
\end{enumerate}

The Bernoulli shift thus yields a simple example for an essentially
nonlinear stretch-and-cut mechanism, as it typically generates {\em
deterministic chaos} \cite{Ott}. Such basic mechanisms are also
encountered in more realistic dynamical systems. We may remark that
`stretch and fold' or `stretch, twist and fold' provide alternative
mechanisms for generating chaotic behavior, see, e.g., the tent
map. The reader may wish to play around with these ideas in thought
experiments, where the sets of points is replaced by kneading
dough. These ideas can be made mathematically precise in form of what
is called {\em mixing} in dynamical systems, which is an important
concept in the ergodic theory of dynamical systems
\cite{ArAv68,Do99}.

\section{Ljapunov chaos} \label{ljap}

In Ref.~\cite{Dev89} Devaney defines chaos by requiring that for a
given dynamical system three conditions have to be fulfilled:
sensitivity, existence of a dense orbit, and that the periodic points
are dense. The {\em Ljapunov exponent} generalizes the concept of
sensitivity in form of a quantity that can be calculated more
conveniently, as we will motivate by an example:

\begin{example} Ljapunov instability of the Bernoulli shift
\cite{Ott,Rob95}

Consider two points that are initially displaced from each other by
$\delta x_0 :=|x_0'-x_0|$ with $\delta x_0$ ``infinitesimally small''
such that $x_0,x_0'$ do not hit different branches of the Bernoulli
shift $B(x)$ around $x=1/2$. We then have
\be
\delta x_n := |x_n'-x_n|=2\delta x_{n-1} = 2^2\delta x_{n-2}
=\ldots=2^n\delta x_0 =e^{n \ln2} \delta x_0 \quad .
\ee
We thus see that there is an {\em exponential separation} between two
nearby points as we follow their trajectories. The rate of separation
$\lambda(x_0):=\ln 2$ is called the (local) {\em Ljapunov exponent} of
the map $B(x)$.

\end{example}

This simple example can be generalized as follows, leading to the
general definition of the Ljapunov exponent for one-dimensional
maps $F$. Consider
\be
\delta x_n = |x_n'-x_n|=|F^n(x_0')-F^n(x_0)|=:\delta x_0
e^{n\lambda(x_0)}\: (\delta x_0\to0)
\ee
for which we {\em presuppose} that an exponential separation
of trajectories exists.\footnote{We emphasize that this is not always
  the case, see, e.g., Section 17.4 of \cite{Kla06}.}  By furthermore
assuming that $F$ is differentiable we can rewrite this equation to
\bna
\lambda(x_0) &=& \lim_{n\to\infty} \lim_{\delta x_0 \to 0}
\frac{1}{n} \ln \frac{\delta x_n}{\delta x_0}\nonumber\\
&=&\lim_{n\to\infty}
\lim_{\delta x_0 \to 0} \frac{1}{n} \ln \frac{\left|F^n(x_0+\delta
x_0)-F^n(x_0)\right|}{\delta x_0} \nonumber \\
&=& \lim_{n\to\infty}\frac{1}{n} \ln\left|\frac{d F^n(x)}{d x}\right|_{x=x_0}\quad .
\ena
Using the chain rule we obtain
\be
\left.\frac{\ud F^n(x)}{\ud x}\right|_{x=x_0}=F'(x_{n-1})
F'(x_{n-2})\ldots F'(x_0) \quad ,
\ee
which leads to
\bna
\lambda(x_0) &=& \lim_{n\to\infty}\frac{1}{n}
\ln\left|\prod_{i=0}^{n-1} F'(x_i)\right| \nonumber \\
&=& \lim_{n\to\infty}\frac{1}{n} \sum_{i=0}^{n-1}
\ln\left|F'(x_i)\right| \quad .
\ena
This simple calculation motivates the following definition:

\begin{defi} \cite{ASY97} \label{def:ljap}
Let $F\in C^1$ be a map of the real line. The {\em local Ljapunov
exponent} $\lambda(x_0)$ is defined as
\be
\lambda(x_0):=\lim_{n\to \infty}\frac{1}{n}\sum_{i=0}^{n-1}\ln \left|F'(x_i)\right|
\ee
if this limit exists.\footnote{This definition was proposed by
  A.M.~Ljapunov in his Ph.D.\ thesis 1892.}
\end{defi}

\begin{remark}
  
\nl
  
\vspace*{-0.7cm}
\begin{enumerate}
\item If $F$ is not $C^1$ but {\em piecewise} $C^1$, the definition
  can still be applied by excluding single points of
  non-differentiability.
\item If $F'(x_i)=0\Rightarrow\not\exists\lambda(x)$. However,
  usually this concerns only an `atypical' set of points.
  \end{enumerate} 
\end{remark}

\begin{example}
For the Bernoulli shift $B(x)=2x \mod 1$ we have $B'(x)=2 \: \forall
x\in [0,1)\:,\: x\neq \half$, hence trivially
\be
\lambda(x)= \frac{1}{n}\sum_{k=0}^{n-1} \ln 2 = \ln 2
\ee
at these points.
\end{example}

Note that Definition~\ref{def:ljap} defines the {\em local} Ljapunov
exponent $\lambda(x_0)$, that is, this quantity may depend on our
choice of initial conditions $x_0$. For the Bernoulli shift this is
not the case, because this map has a uniform slope of two except at
the point of discontinuity, which makes the calculation trivial.
Generally the situation is more complicated.  One question is then of
how to calculate the local Ljapunov exponent, and another one to which
extent it depends on initial conditions. An answer to both these
questions is provided in form of the {\em global} Ljapunov exponent
that we are going to introduce, which does not depend on initial
conditions and thus characterizes the stability of the map as a whole.

It is introduced by observing that the local Ljapunov exponent in
Definition~\ref{def:ljap} is defined in form of a {\em time average},
where $n$ terms along the trajectory with initial condition $x_0$ are
summed up by averaging over $n$. That this is not the only possibility
to define an average quantity is clarified by the following
definition:

\begin{defi} time and ensemble average \cite{Do99,ArAv68}

Let $\mu^*$ be the invariant probability measure of a one-dimensional
map $F$ acting on $J\subseteq \mathbb{R}$. Let us consider a function
$g:J\to \mathbb{R}$, which we may call an ``observable''. Then
\be
\overline{g(x)} := \lim_{n\to\infty} \frac{1}{n}\sum_{k=0}^{n-1} g(x_k) \label{eq:tiav}\quad ,
\ee 
$x=x_0$, is called the {\em time (or Birkhoff) average} of $g$ with respect to
$F$.
\be
\langle g\rangle := \int_J d\mu^* g(x) \label{eq:enav}
\ee
where, if such a measure exists, $d\mu^*=\rho^*(x)\:dx$, is called the
{\em ensemble (or space) average} of $g$ with respect to $F$.  Here
$\rho^*(x)$ is the {\em invariant density} of the map, and $d\mu^*$ is
the associated {\em invariant measure} \cite{Ott,LM94,dslnotes}. Note
that $\overline{g(x)}$ may depend on $x$, whereas $\langle g\rangle$
does not.  \end{defi}

If we choose $g(x)=\ln|F'(x)|$ as the observable in
Eq.~(\ref{eq:tiav}) we recover Definition~\ref{def:ljap} for the local
Ljapunov exponent,
\be
\lambda(x):=\overline{\ln|F'(x)|}=\lim_{n\to\infty} \frac{1}{n}\sum_{k=0}^{n-1} 
\ln|F'(x_k)|\quad ,
\ee
which we may write as $\lambda_t(x)=\lambda(x)$ in order to
label it as a time average. If
we choose the same observable for the ensemble average
Eq.~(\ref{eq:enav}) we obtain
\be
\lambda_e:=\langle\ln|F'(x)|\rangle := \int_J dx{\rho^*(x)} \ln|F'(x)| \quad . \label{eq:ljapens}
\ee

\begin{example} \label{ex:ljap}

For the Bernoulli shift we have seen that for almost every $x\in[0,1)$
$\lambda_t=\ln 2$. For $\lambda_e$ we obtain
\be
\lambda_e=\int_0^1 dx \rho^*(x) \ln 2 = \ln 2 \quad ,
\ee
taking into account that $\rho^*(x)=1$, as you have seen before.  
In other words, time and
ensemble average are the same for almost every $x$,
\be
\lambda_t(x)=\lambda_e=\ln 2 \quad .
\ee
\end{example}

This motivates the following fundamental definition:
\begin{defi} ergodicity \label{def:ergo} \cite{ArAv68,Do99}\footnote{Note that mathematicians prefer to define ergodicity by using the concept of indecomposability \cite{KaHa95}.}

A dynamical system is called {\em ergodic} if for every $g$ on
$J\subseteq \mathbb{R}$ satisfying $\int
d\mu^*\:|g(x)|<\infty$\footnote{This condition means that we require
$g$ to be a Lebesgue-integrable function. In other words, $g$ should
be an element of the function space $L^1(J,\mathscr{A},\mu^*)$ of a
set $J$, a $\sigma$-algebra $\mathscr{A}$ of subsets of $J$ and an
invariant measure $\mu^*$.  This space defines the family of all
possible real-valued measurable functions $g$ satisfying $\int
d\mu^*\:|g(x)|<\infty$, where this integral should be understood as a
Lebesgue integral \cite{LM94}.}
\be 
\overline{g(x)}=\langle g\rangle 
\ee 
for typical $x$.

\end{defi} 

For our purpose it suffices to think of a typical $x$ as a point that
is randomly drawn from the invariant density $\rho^*(x)$.  This
definition implies that for ergodic dynamical systems
$\overline{g(x)}$ does not depend on $x$. That the time average is
constant is sometimes also taken as a definition of ergodicity
\cite{Do99,Beck}. To prove that a given system is ergodic is typically
a hard task and one of the fundamental problems in the ergodic theory
of dynamical systems; see \cite{ArAv68,Do99,TKS92} for proofs of
ergodicity in case of some simple examples.

On this basis, let us get back to Ljapunov exponents.  For time
average $\lambda_t(x)$ and ensemble average $\lambda_e$ of the
Bernoulli shift we have found that $\lambda_t(x)=\lambda_e=\ln 2$.
Definition~\ref{def:ergo} now states that the first equality must hold
whenever a map $F$ is ergodic. This means, in turn, that for an
ergodic dynamical system the Ljapunov exponent becomes a {\em global}
quantity characterizing a given map $F$ for a typical point $x$
irrespective of what value we choose for the initial condition,
$\lambda_t(x)=\lambda_e=\lambda$.  This observation very much
facilitates the calculation of $\lambda$, as is demonstrated by the
following example:

\begin{example}

Let us consider the map $A(x)$ displayed in Fig.~\ref{fig:mapfljap} below:
  
\begin{figure}[htp]
\begin{center}
\includegraphics[height=5cm,angle=-90]{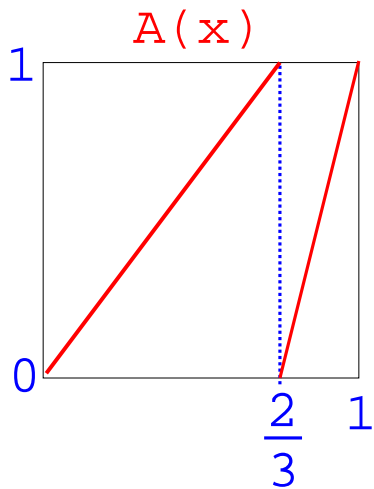}
\caption{A simple map for demonstrating the calculation of
Ljapunov exponents via ensemble averages.}  
\label{fig:mapfljap}
\end{center}
\end{figure}

From the figure we can infer that 
\be
A(x):= \left\{\begin{array}{l} \frac{3}{2}x\;,\; 0\le x < \frac{2}{3} \\ 
3x-2\;,\; \frac{2}{3} \le x < 1 \end{array}\right. \quad . \label{eq:mapa}
\ee
It is not hard to see that the invariant probability density of
this map is uniform, $\rho^*(x)=1$. The Ljapunov exponent $\lambda$
for this map is then trivially calculated to
\be
\lambda = \int_0^1\ud x \rho^*(x) \ln|A'(x)| = 
 \ln 3 - \frac{2}{3}\ln 2\quad .
\ee
By assuming that map $A$ is ergodic (which here is the case), we
can conclude that this result for $\lambda$ represents the value for
typical points in the domain of $A$.

\end{example}

In other words, for an ergodic map the global Ljapunov exponent
$\lambda$ yields a number that assesses whether it is chaotic in the
sense of exhibiting an exponential dynamical instability. This
motivates the following definition of deterministic chaos:

\begin{defi} Chaos in the sense of Ljapunov \cite{Rob95,Ott,ASY97,Beck}  

  An ergodic map $F: J\to J, \: J\subseteq \mathbb{R}, \: F$ (piecewise) $C^1$
  is said to be {\em L-chaotic} on $J$ if $\lambda>0$.
\end{defi}

Why did we introduce a definition of chaos that is different from
Devaney's definition mentioned earlier? One reason is that often the
largest Ljapunov exponent of a dynamical system is easier to calculate
than checking for sensitivity.\footnote{Note that a positive Ljapunov
exponent implies sensitivity, but the converse does not hold true
\cite{dslnotes}.}  Furthermore, the magnitude of the positive Ljapunov
exponent quantifies the strength of chaos. This is the reason why in
the applied sciences ``chaos in the sense of Ljapunov'' became a very
popular concept.\footnote{Here one often silently assumes that a given
dynamical system is ergodic. To prove that a system is topologically
transitive as required by Devaney's definition is not any easier.} 
Note that there is no unique quantifier of deterministic chaos. Many
different definitions are available highlighting different aspects of
``chaotic behavior'', all having their advantages and
disadvantages. The detailed relations between them (such as the ones
between Ljapunov and Devaney chaos) are usually non-trivial and a
topic of ongoing research \cite{dslnotes}.  We will encounter yet
another definition of chaos in the following section.

\section{Entropies}\label{sec:entropy}

This section particularly builds upon the presentations in
\cite{Ott,Do99}; for a more mathematical approach see \cite{ER}. Let
us start with a brief motivation outlining the basic idea of entropy
production in dynamical systems. Consider again the Bernoulli shift by
decomposing its domain $J=[0,1)$ into $J_0:=[0,1/2)$ and
$J_1:=[1/2,1)$.  For $x\in[0,1)$ define the {\em output map} $s$ by
\cite{Schu}
\bna
s:[0,1)\to\{0,1\}\;,\;s(x):= \left\{\begin{array}{l} 0\;,\;  x\in J_0\\ 
1\;,\; x\in J_1 \end{array}\right. \label{eq:benc}
\ena 
and let $s_{n+1}:=s(x_n)$. Now choose some initial condition $x_0\in
J$. According to the above rule we obtain a digit $s_1\in\{0,1\}$.
Iterating the Bernoulli shift according to $x_{n+1}=B(x_n)$ then
generates a sequence of digits $\{s_1,s_2,\ldots,s_n\}$. This sequence
yields nothing else than the binary representation of the given
initial condition $x_0$ \cite{Schu,Ott,Do99}. If we assume that we
pick an initial condition $x_0$ at random and feed it into our map
without knowing about its precise value, this simple algorithm enables
us to find out what number we have actually chosen.  In other words,
here we have a mechanism of {\em creation of information} about the
initial condition $x_0$ by analyzing the chaotic orbit generated from
it as time evolves.

Conversely, if we now assume that we already knew the initial state up
to, say, $m$ digits precision and we iterate $p>m$ times, we see that
the map simultaneously {\em destroys information} about the current
and future states, in terms of digits, as time evolves. So creation of
information about previous states goes along with loss of information
about current and future states. This process is quantified by the
{\em Kolmogorov-Sinai (KS) entropy} (also called metric, or
measure-theoretic entropy), which measures the exponential rate at
which information is produced, respectively lost in a dynamical
system, as we will see below.

\begin{figure}[t]
\begin{center}
\includegraphics[width=4cm,angle=-90]{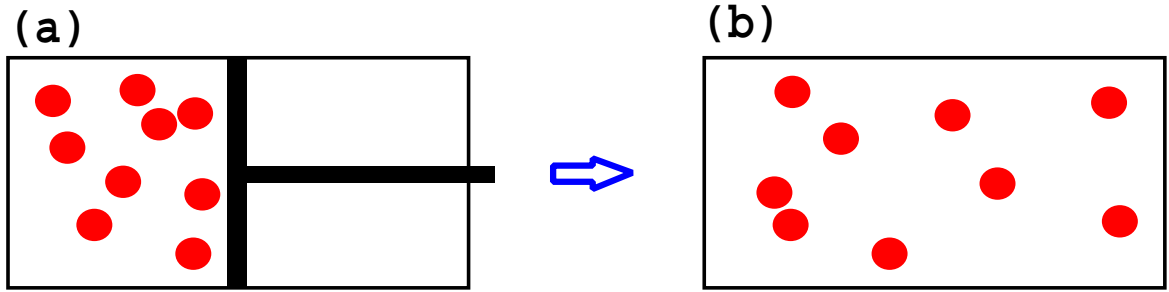}
\caption{Schematic representation of a gas of molecules in a box. In (a) 
  the gas is constrained by a piston to the left hand side of the box,
  in (b) the piston is removed and the gas can spread out over the
  whole box.  This illustrates the basic idea of (physical) entropy
  production.}
\label{fig:phep}
\end{center}
\end{figure}

The situation is similar to the following thought experiment
illustrated in Fig.~\ref{fig:phep}: Let us assume we have a gas
consisting of molecules, depicted as billiard balls, which is
constrained to the left half of the box as shown in (a). This is like
having some information about the initial conditions of all gas
molecules, which are in a more localized, or ordered, state.  If we
remove the piston as in (b), we observe that the gas spreads out over
the full box until it reaches a uniform equilibrium steady state. We
then have less information available about the actual positions of all
gas molecules, that is, we have increased the disorder of the whole
system. This observation lies at the heart of what is called {\em
thermodynamic entropy production} in the statistical physics of
many-particle systems which, however, is usually assessed by
quantities that are different from the KS-entropy \cite{FPV05}.

At this point we may not further elaborate on the relation to
statistical physical theories. Instead, let us make precise what we
mean by KS-entropy starting from the famous {\em Shannon (or
information) entropy} \cite{Ott,Beck}. This entropy is defined as
\be
H_S:=\sum_{i=1}^r p_i\ln\left(\frac{1}{p_i}\right) \quad , \label{eq:shannon}
\ee
where $p_i\:,\:i=1,\ldots,r$ are the probabilities for the $r$
possible outcomes of an experiment. Think, for example, of a roulette
game, where carrying out the experiment one time corresponds to $n=1$
in the iteration of an unknown map. $H_S$ then measures the amount of
uncertainty concerning the outcome of the experiment, which can be
understood as follows: 
\begin{enumerate}
\item Let $p_1=1\:,\:p_i=0$ otherwise. By defining
  $p_i\ln\left(\frac{1}{p_i}\right):=0\:,\:i\neq 1,$ we have
  $H_S=0$. This value of the Shannon entropy must therefore
  characterize the situation where the outcome is completely certain.
\item Let $p_i=1/r\:,\:i=1,2,\ldots,r$. Then we obtain $H_S=\ln r$
  thus characterizing the situation where the outcome is most
  uncertain because of equal probabilities.
\end{enumerate}

Case (1) thus represents the situation of no information gain by doing
the experiment, case (2) corresponds to maximum information gain.
These two special cases must therefore define the lower and upper
bounds of $H_S$,\footnote{A proof employs the convexity of the entropy
function and Jensen's inequality, see \cite{BaPo97} or the Wikipedia
entry of {\em information entropy} for details.}
\be
0\le H_S\le \ln r \quad .
\ee
This basic concept of information theory carries over to dynamical
systems by identifying the probabilities $p_i$ with invariant
probability measures $\mu_i^*$ on subintervals of a given dynamical
system's phase space. The precise connection is worked out in four
steps \cite{Ott,Do99}:\\

\noindent {\bf 1. Partition and refinement:}

Consider a map $F$ acting on $J\subseteq \mathbb{R}$, and let $\mu^*$
be an invariant probability measure generated by the map.\footnote{If
  not said otherwise, $\mu^*$ holds for the {\em physical (or natural)
    measure} of the map in the following \cite{dslnotes}.} Let
$\{J_i\},\ i=1,\ldots, s$ be a {\em partition} of $J$.\footnote{A {\em
    partition} of the interval $J$ is a collection of subintervals
  whose union is $J$, which are pairwise disjoint except perhaps at
  the end points \cite{ASY97}.} We now construct a {\em refinement} of
this partition as illustrated by the following example:

\begin{example}
  
  Consider the Bernoulli shift displayed in Fig.~\ref{fig:bpart}.
  Start with the partition $\{J_0,J_1\}$ shown in (a).  Now create
  a {\em refined} partition by iterating these two partition parts
  backwards according to $B^{-1}(J_i)$ as indicated in (b).  
  Alternatively, you may take the second forward iterate $B^2(x)$
  of the Bernoulli shift and then identify the preimages of $x=1/2$
  for this map. In either case the new partition parts are 
  obtained to
\bna
J_{00}&:=&\{x:x\in J_0\:,\:B(x)\in J_0\}\nonumber \\
J_{01}&:=&\{x:x\in J_0\:,\:B(x)\in J_1\}\nonumber \\
J_{10}&:=&\{x:x\in J_1\:,\:B(x)\in J_0\}\nonumber \\
J_{11}&:=&\{x:x\in J_1\:,\:B(x)\in J_1\} \quad .
\ena

\begin{figure}[htp]
\begin{center}
\includegraphics[width=8cm,angle=-90]{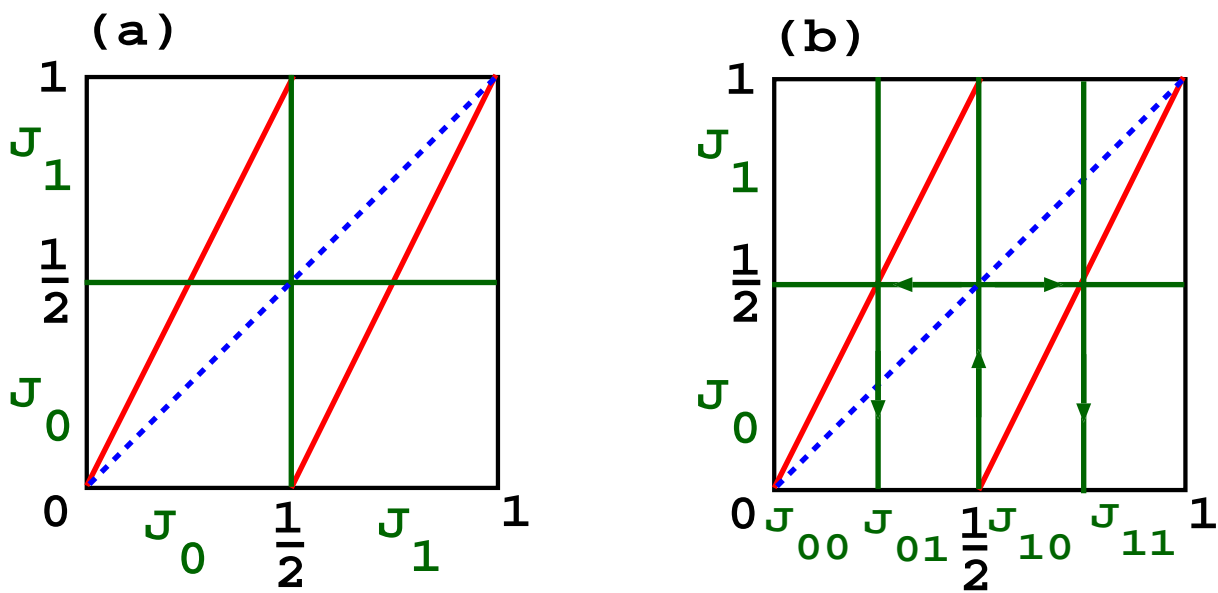}
\caption{(a) The Bernoulli shift and a partition of the unit interval 
  consisting of two parts. (b) Refinement of this partition under
  backward iteration.}
\label{fig:bpart}
\end{center}
\end{figure}
If we choose $x_0\in J_{00}$ we thus know in advance that the orbit
emerging from this initial condition under iteration of the map will
remain in $J_0$ at the next iteration. That way, the refined partition
clearly yields more information about the dynamics of single orbits.
\end{example}

More generally, for a given map $F$ the above procedure is equivalent to defining
\be
\{J_{i_1i_2}\}:=\{J_{i_1}\cap F^{-1}(J_{i_2})\}\quad .
\ee
The next round of refinement proceeds along the same lines yielding
\be
\{J_{i_1i_2i_3}\}:=\{J_{i_1}\cap F^{-1}(J_{i_2})\cap F^{-2}(J_{i_3})\}\quad ,
\ee
and so on. For convenience we define
\be
\{J_i^n\}:=\{J_{i_1i_2\ldots i_n}\}=\{J_{i_1}\cap F^{-1}(J_{i_2})\cap \ldots\cap F^{-(n-1)}(J_{i_n})\}\quad .
\ee

\noindent {\bf 2. H-function:}

In analogy to the Shannon entropy Eq.~(\ref{eq:shannon}) next we define the function
\be
H(\{J_i^n\}):=-\sum_i\mu^*(J_i^n)\ln\mu^*(J_i^n)\quad , \label{eq:hndef}
\ee
where $\mu^*(J_i^n)$ is the invariant measure of the map $F$ on the
partition part $J_i^n$ of the $n$th refinement.

\begin{example}\label{ex:hksb}
  
  For the Bernoulli shift with uniform invariant probability density
  $\rho^*(x)=1$ and associated (Lebesgue) measure $\mu^*(J_i^n)=\int_
  {J_i^n} dx \:\rho^*(x)=\mbox{diam} \left(J_i^n\right)$ we can calculate
\bna
H(\{J_i^1\})&=&-(\frac{1}{2}\ln\frac{1}{2}+\frac{1}{2}\ln\frac{1}{2})=\ln 2\nonumber\\
H(\{J_i^2\})&=&H(\{J_{i_1}\cap B^{-1}(J_{i_2})\})=-4(\frac{1}{4}\ln\frac{1}{4})=\ln 4\nonumber\\
H(\{J_i^3\})&=&\ldots =\ln 8=\ln 2^3\nonumber\\
 &\vdots&\nonumber\\
H(\{J_i^n\})&=&\ln 2^n
\ena

\end{example}

\noindent {\bf 3. Take the limit:}

We now look at what we obtain in the limit of infinitely refined partition by
\be
h(\{J_i^n\}):=\lim_{n\to\infty}\frac{1}{n} H(\{J_i^n\}) \quad , \label{eq:hlim}
\ee
which defines the rate of gain of information over $n$ refinements.

\begin{example}

For the Bernoulli shift we trivially obtain\footnote{In fact, this
result is already obtained after a single iteration step, i.e.,
without taking the limit of $n\to\infty$. This reflects the fact that
the Bernoulli shift dynamics sampled that way is mapped onto a
Markov process.}
\be
h(\{J_i^n\})=\ln 2 \quad .
\ee

\end{example}

\noindent {\bf 4. Supremum over partitions:}

We finish the definition of the KS-entropy by maximizing $h(\{J_i^n\})$
over all available partitions,
\be
h_{KS}:=\sup_{\{J_i^n\}} h(\{J_i^n\}) \quad .
\ee
The last step can be avoided if the partition $\{J_i^n\}$ is {\em
  generating} for which it must hold that $\mbox{diam}
\left(J_i^n\right)\to 0\;(n\to\infty)$ \cite{ER,KaHa95,BaPo97}.\footnote{Note
  that alternative definitions of a generating partition in terms of
  symbolic dynamics are possible \cite{Beck}.} It is quite obvious
that for the Bernoulli shift the partition chosen above is generating
in that sense, hence $h_{KS}=\ln 2$ for this map.

\begin{remark} \cite{Schu}
  
  
\begin{enumerate}
\item For strictly periodic motion there is no refinement of partition
  parts under backward iteration, hence $h_{KS}=0$; see,
  for example, the identity map $I(x)=x$.
\item For stochastic systems all preimages are possible, so there is
  immediately an infinite refinement of partition parts under backward
  iteration, which leads to $h_{KS}\to\infty$.
\end{enumerate}

\end{remark}

These considerations suggest yet another definition of deterministic
chaos:

\begin{defi} Measure-theoretic chaos \cite{Beck}  
  
  A map $F: J\to J, \: J\subseteq \mathbb{R},$ is said to be chaotic
  in the sense of exhibiting {\em dynamical randomness} if $h_{KS}>0$.
\end{defi}

Again, one may wonder about the relation between this new definition
and our previous one in terms of Ljapunov chaos. Let us look again at
the Bernoulli shift:

\begin{example}

For $B(x)$ we have calculated the Ljapunov exponent to $\lambda=\ln
2$, see Example~\ref{ex:ljap}. Above we have seen that $h_{KS}=\ln 2$
for this map, so we arrive at $\lambda=h_{KS}=\ln 2$.

\end{example}

That this equality is not an artefact due to the simplicity of our
chosen model is stated by the following theorem:

 \begin{theo} Pesin's Theorem (1977) \cite{Do99,ER,Young02}
   
   For {\em closed $C^2$ Anosov\footnote{An
   Anosov system is a diffeomorphism, where the expanding and
   contracting directions in phase space exhibit a particularly
   ``nice'', so-called hyperbolic structure \cite{ER,Do99}.} systems}
   the KS-entropy is equal to the sum of positive Ljapunov
   exponents.\\

  \end{theo}
  
  A proof of this theorem goes considerably beyond the scope of this
  course \cite{Young02}. In the given formulation it applies to
  higher-dimensional dynamical systems that are ``suitably
  well-behaved'' in the sense of exhibiting the Anosov property.
  Applied to one-dimensional maps, it means that if we consider
  transformations which are ``closed'' by mapping an interval onto
  itself, $F:\:J\to J$, under certain conditions (which we do not
  further specify here) and if there is a positive Ljapunov exponent
  $\lambda>0$ we can expect that $\lambda=h_{KS}$, as we have seen for
  the Bernoulli shift. In fact, the Bernoulli shift provides an
  example of a map that does not fulfill the conditions of the above
  theorem precisely. However, the theorem can also be formulated under
  weaker assumptions, and it is believed to hold for an even wider
  class of dynamical systems.
  
  In order to get an intuition why this theorem should hold, let us
  look at the information creation in a simple one-dimensional map
  such as the Bernoulli shift by considering two orbits
  $\{x_k\}_{k=0}^n$, $\{x'_k\}_{k=0}^n$ starting at nearby initial
  conditions $|x_0'-x_0|\le\delta x_0\:,\:\delta x_0\ll 1$. Recall the
  encoding defined by Eq.~(\ref{eq:benc}). Under the first $m$
  iterations these two orbits will then produce the very same
  sequences of symbols $\{s_k\}_{k=1}^m$, $\{s'_k\}_{k=1}^m$, that is,
  we cannot distinguish them from each other by our encoding. However,
  due to the ongoing stretching of the initial displacement $\delta
  x_0$ by a factor of two, eventually there will be an $m$ such that
  starting from $p>m$ iterations different symbol sequences are
  generated. Thus we can be sure that in the limit of $n\to\infty$ we
  will be able to distinguish initially arbitrarily close orbits. If
  you like analogies, you may think of extracting information about
  the different initial states via the stretching produced by the
  iteration process like using a magnifying glass. Therefore, under
  iteration the exponential rate of separation of nearby trajectories,
  which is quantified by the positive Ljapunov exponent, must be equal
  to the rate of information generated, which in turn is given by the
  KS-entropy. This is at the heart of Pesin's theorem.

We may remark that typically the KS-entropy is much harder to
calculate for a given dynamical system than positive Ljapunov
exponents. Hence, Pesin's theorem is often employed in the literature
for indirectly calculating the KS-entropy. Furthermore, here we have
described only one type of entropy for dynamical systems. It should be
noted that the concept of the KS-entropy can straightforwardly be
generalized leading to a whole spectrum of {\em Renyi entropies},
which can then be identified with topological, metric, correlation and
other higher-order entropies \cite{Beck}.

\section{Open systems, fractals and escape rates}

This section draws particularly on \cite{Ott,Do99}. So far we have
only studied closed systems, where intervals are mapped onto
themselves. Let us now consider an {\em open system}, where points can
leave the unit interval by never coming back to it. Consequently, in
contrast to closed systems the total number of points is not conserved
anymore. This situation can be modeled by a slightly generalized
example of the Bernoulli shift.

\begin{example}\label{ex:obern}
In the following we will study the map
\be
B_a:[0,1)\to[1-a/2,a/2)\;,\;B_a(x):= \left\{\begin{array}{l} ax\;,\; 0\le x < 1/2 \\ 
ax+1-a\;,\; 1/2 \le x < 1 \end{array}\right. \quad , \label{eq:obern}
\ee
see Fig.~\ref{fig:obern}, where the slope $a\ge 2$ defines a control
parameter. For $a=2$ we recover our familiar Bernoulli shift, whereas
for $a>2$ the map defines an open system. That is, whenever points
are mapped into the escape region of width $\Delta$ these points are
removed from the unit interval. You may thus think of the escape
region as a subinterval that absorbs any particles mapped onto it.

\begin{figure}[htp]
\begin{center}
\includegraphics[width=8cm,angle=-90]{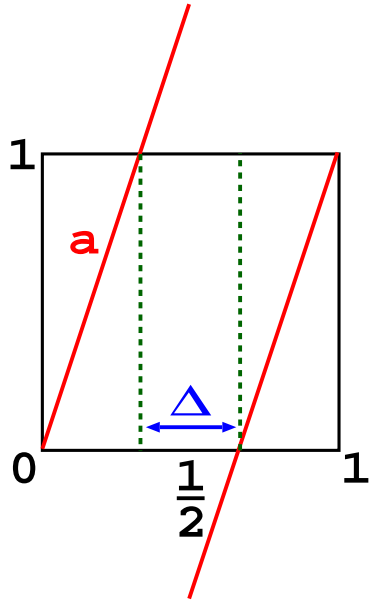}
\caption{A generalization of the Bernoulli shift, defined as a
  parameter-dependent map $B_a(x)$ modeling an open system. The slope
  $a$ defines a control parameter, $\Delta$ denotes the width of the
  escape region.}
\label{fig:obern} 
\end{center} 
\end{figure}

\end{example}

We now wish to compute the number of points $N_n$ remaining on the
unit interval at time step $n$, where we start from a uniform
distribution of $N_0=N$ points on this interval at $n=0$. This can be
done as follows: Recall that the probability density $\rho_n(x)$ was
defined by
\be\displaystyle
\rho_n(x):=\frac{\mbox{number of points} \:N_{n,j} \:\mbox{in interval 
dx centered around position} \: x_j \:\mbox{at time step}\:
n}{\mbox{total number of points}\: N\:\mbox{times width dx} }\quad , \label{eq:pdfdef}
\ee
where $N_n=\sum_j N_{n,j}$. With this we have that
\be
N_1=N_0-\rho_0 N\Delta \quad .
\ee
By observing that for $B_a(x)$, starting from $\rho_0=1$ points are
always uniformly distributed on the unit interval at subsequent
iterations, we can derive an equation for the density $\rho_1$ of
points covering the unit interval at the next time step $n=1$. For
this purpose we divide the above equation by the total number of
points $N$ (multiplied with the total width of the unit interval,
which however is one), which yields
\be
\rho_1=\frac{N_1}{N}=\rho_0-\rho_0\Delta=\rho_0(1-\Delta) \quad .
\ee
This procedure can be reiterated starting now from
\be
N_2=N_1-\rho_1 N\Delta 
\ee
leading to
\be
\rho_2=\frac{N_2}{N}=\rho_1(1-\Delta)\quad ,
\ee
and so on. For general $n$ we thus obtain
\be
\rho_n=\rho_{n-1}(1-\Delta)=\rho_0(1-\Delta)^n=\rho_0 e^{n\ln (1-\Delta)}\quad ,
\ee
or correspondingly
\be
N_n=N_0 e^{n\ln (1-\Delta)}\quad ,
\ee
which suggests the following definition:

\begin{defi}
For an open system with exponential decrease of the number of points,
\be
N_n=N_0 e^{-\gamma n}\quad , \label{eq:partdec}
\ee
$\gamma$ is called the {\em escape rate}.
\end{defi}

\begin{figure}[t]
\begin{center}
\includegraphics[width=8cm]{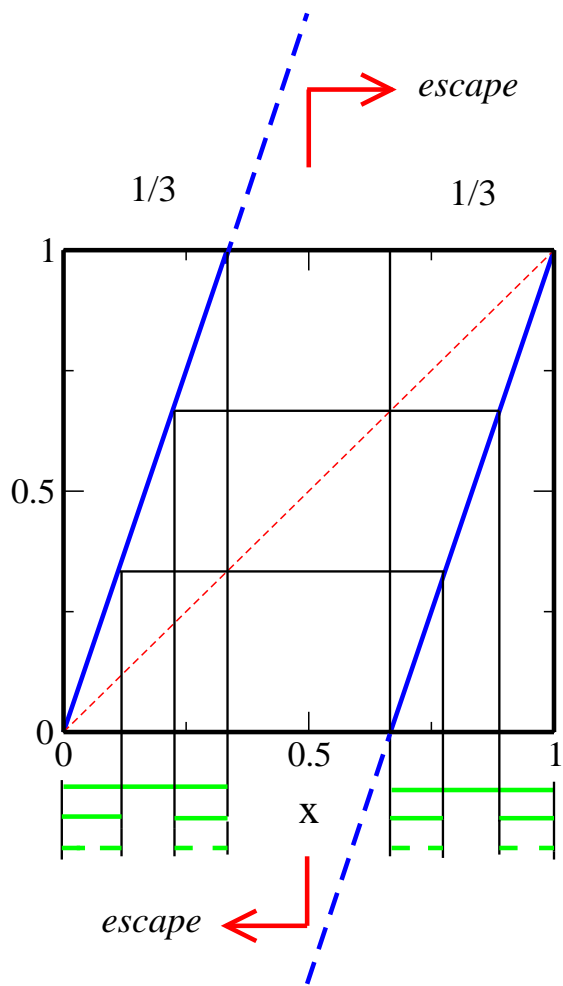}
\caption{Construction of the set ${\cal C}_{B_3}$ of initial conditions of the map $B_3(x)$ that never leave the unit interval.}  
\label{fig:bcantor} 
\end{center} 
\end{figure}

In case of our mapping we thus identify
\be
\gamma=\ln\frac{1}{1-\Delta} \label{eq:escape}
\ee
as the escape rate. We may now wonder whether there are any initial
conditions that never leave the unit interval and about the character
of this set of points. The set can be constructed as exemplified for
$B_a(x)\:,\:a=3$ in Fig.~\ref{fig:bcantor}.

\begin{example}

Let us start again with a uniform distribution of points on the unit
interval. We can then see that the points which remain on the unit
interval after one iteration of the map form two sets, each of length
$1/3$. Iterating now the boundary points of the escape region
backwards in time according to $x_n=B_3^{-1}(x_{n+1})$, we can obtain
all preimages of the escape region. We find that initial points which
remain on the unit interval after two iterations belong to four
smaller sets, each of length $1/9$, as depicted at the bottom of
Fig.~\ref{fig:bcantor}. Repeating this procedure infinitely many times
reveals that the points which never leave the unit interval form the
very special set ${\cal C}_{B_3}$, which is known as the {\em middle
third Cantor set}.

\end{example}

\begin{defi} Cantor set \cite{Ott,Dev89}

A {\em Cantor set} is a closed set which consists entirely of boundary
points each of which is a limit point of the set.
\end{defi}

Let us explore some fundamental properties of the set ${\cal C}_{B_3}$
\cite{Ott}:

\begin{enumerate}
  
\item From Fig.~\ref{fig:bcantor} we can infer that the total length
  $l_n$ of the intervals of points remaining on the unit interval
  after $n$ iterations, which is identical with the Lebesgue measure
  $\mu_L$ of these sets, is
\be
l_0=1\:,\:l_1=\frac{2}{3}\:,\:l_2=\frac{4}{9}=\left(\frac{2}{3}\right)^2\:,\ldots,l_n=\left(\frac{2}{3}\right)^n \quad .
\ee
We thus see that 
\be
l_n=\left(\frac{2}{3}\right)^n \to 0\:(n\to\infty) \quad ,
\ee
that is, the total length of this set goes to zero, $\mu_L({\cal
C}_{B_3})=0$. However, there exist also Cantor sets whose
Lebesgue measure is larger than zero \cite{Ott}.  Note that
matching $l_n=\exp(-n\ln(3/2))$ to Eq.~(\ref{eq:escape})
yields an escape rate of $\gamma=\ln (3/2)$ for this map.

\item By using the binary encoding Eq.~(\ref{eq:benc}) for all
intervals of ${\cal C}_{B_3}$, thus mapping all elements of this set
onto all the numbers in the unit interval, it can nevertheless be
shown that our Cantor set contains an {\em uncountable number of
points} \cite{Do99,Tric95}.

\item By construction ${\cal C}_{B_3}$ must be the invariant set of
the map $B_3(x)$ under iteration, so the invariant measure of our open
system must be the measure defined on the Cantor set, $\mu^*({\cal
C})\:,\:{\cal C}\in {\cal C}_{B_3}$ \cite{KaHa95}; see the following
Example \ref{ex:escape} for the procedure of how to calculate this
measure.

\item For the next property we need the following definition:

\begin{defi} repeller \cite{Beck,Do99}

The limit set of points that never escape is called a {\em
repeller}. The orbits that escape are {\em transients}, and $1/\gamma$
is the typical duration of them.

\end{defi}

From this we can conclude that ${\cal C}_{B_3}$ represents the
repeller of the map $B_3(x)$.

\item Since ${\cal C}_{B_3}$ is completely disconnected by only
consisting of boundary points, its topology is highly
singular. Consequently, no invariant density $\rho^*(x)$ can be
defined on this set, since this concept presupposes a certain
``smoothness'' of the underlying topology such that one can
meaningfully speak of ``small subintervals dx'' on which one counts
the number of points, see Eq.~(\ref{eq:pdfdef}). In contrast,
$\mu^*({\cal C})$ is still well-defined,\footnote{This is one of the
reasons why mathematicians prefer to deal with measures instead of
densities.} and we speak of it as a {\em singular measure}
\cite{dslnotes,Do99}.

\item Fig.~\ref{fig:bcantor} shows that ${\cal C}_{B_3}$ is {\em
    self-similar}, in the sense that smaller pieces of this structure
  reproduce the entire set upon magnification \cite{Ott}. Here we find
  that the whole set can be reproduced by magnifying the fundamental
  structure of two subsets with a gap in the middle by a constant
  factor of three. Often such a simple scaling law does not exist for
  these types of sets. Instead, the scaling may depend on the position
  $x$ of the subset, in which case one speaks of a {\em self-affine}
  structure \cite{Man82,Falc90}.

\item Again we need a definition:

\begin{defi} fractals, qualitatively \cite{Beck,Falc90}

{\em Fractals} are geometrical objects that possess nontrivial
structure on arbitrarily fine scales.

\end{defi}

In case of our Cantor set ${\cal C}_{B_3}$, these structures are
generated by a simple scaling law. However, generally fractals can be
arbitrarily complicated on finer and finer scales. A famous example of
a fractal in nature, mentioned in the pioneering book by Mandelbrot
\cite{Man82}, is the coastline of Britain. An example of a structure
that is trivial, hence not fractal, is a straight line. The fractality
of such complicated sets can be assessed by quantities called {\em
fractal dimensions} \cite{Ott, Beck}, which generalize the integer
dimensionality of Euclidean geometry. It is interesting how in our
case fractal geometry naturally comes into play, forming an important
ingredient of the theory of dynamical systems. However, here we do not
further elaborate on the concept of fractal geometry and refer to the
literature instead \cite{Tric95,Falc90,Man82,Ott}.

\end{enumerate}

\begin{example} \label{ex:escape}
  
  Let us now compute all three basic quantities that we have
  introduced so far, that is: the Ljapunov exponent $\lambda$ and the
  KS-entropy $h_{ks}$ {\em on} the invariant set as well as the escape
  rate $\gamma$ {\em from} this set. We do so for the map $B_3(x)$
  which, as we have learned, produces a fractal repeller.  According
  to Eqs.~(\ref{eq:enav}),(\ref{eq:ljapens}) we have to calculate
\be
\lambda({\cal C}_{B_3})=\int_0^1 d\mu^* \ln|B_3'(x)| \quad .
\ee
However, for typical points we have $B_3'(x)= 3$, hence
the Ljapunov exponent must trivially be
\be
\lambda({\cal C}_{B_3})=\ln 3 \quad ,
\ee
because the probability measure $\mu^*$ is normalised. 
The calculation of the KS-entropy requires a bit more work: Recall that
\be
H(\{{\cal C}_i^n\}):=-\sum_{i=1}^{2^n}\mu^*({\cal C}_i^n)\ln\mu^*({\cal C}_i^n)\quad ,
\ee
see Eq.~(\ref{eq:hndef}), where ${\cal C}_i^n$ denotes the $i$th
part of the emerging Cantor set at the $n$th level of its
construction. We now proceed along the lines of
Example~\ref{ex:hksb}. From Fig.~\ref{fig:bcantor} we can infer that
\bna
\mu^*({\cal C}_i^1) &=&\frac{\:\frac{1}{3}\:}{\:\frac{2}{3}\:}=\frac{1}{2}\nonumber \quad 
\ena
at the first level of refinement. Note that here we have {\em
renormalized} the (Lebesgue) measure on the partition part ${\cal
C}_i^1$. That is, we have divided the measure by the total measure
surviving on all partition parts such that we always arrive at a
proper probability measure under iteration. The measure constructed
that way is known as the {\em conditionally invariant measure} on the
Cantor set \cite{Tel90,Beck}. Repeating this procedure yields
\bna
\mu^*({\cal C}_i^2) &=&\frac{\:\frac{1}{9}\:}{\frac{4}{9}}=\frac{1}{4}\nonumber\\
 &\vdots&\nonumber\\
\mu^*({\cal C}_i^n) &=&\frac{(\frac{1}{3})^n}{(\frac{2}{3})^n}=2^{-n}
\ena
from which we obtain
\be
H(\{{\cal C}_i^n\})=-\sum_{i=1}^{2^n}2^{-n}\ln 2^{-n}=n\ln 2\quad .
\ee
We thus see that by taking the limit according to Eq.~(\ref{eq:hlim}) and noting that our partitioning is generating on the fractal repeller ${\cal C}_{B_3}=\{{\cal C}_i^{\infty}\}$, we arrive at
\be
h_{KS}({\cal C}_{B_3})=\lim_{n\to\infty}\frac{1}{n}H(\{{\cal C}_i^n\})=\ln 2 \quad .
\ee
Finally, with Eq.(\ref{eq:escape}) and an escape region of size $\Delta=1/3$ for $B_3(x)$ we get for the escape rate
\be
\gamma({\cal C}_{B_3})=\ln\frac{1}{1-\Delta}=\ln\frac{3}{2} \quad , 
\ee
as we have already seen before.\footnote{Note that the escape rate will generally depend not only on the size but also on the position of the escape interval \cite{BuYu08}.}
\end{example}

In summary, we have that $\gamma({\cal
C}_{B_3})=\ln\frac{3}{2}=\ln 3-\ln 2$, $\lambda({\cal C}_{B_3})=\ln 3$,
$h_{KS}({\cal C}_{B_3})=\ln 2$, which suggests the relation
\be
\gamma({\cal C}_{B_3})=\lambda({\cal C}_{B_3})-h_{KS}({\cal C}_{B_3}) \quad .
\label{eq:escrf}
\ee
Again, this equation is no coincidence. It is a generalization of
Pesin's theorem to open systems, known as the {\em escape rate
formula} \cite{KaGr85}. This equation holds under similar conditions
like Pesin's theorem, which is recovered from it if there is no escape
\cite{Do99}.

\chapter{Deterministic diffusion}\label{chap:detdif}

We now apply the concepts of dynamical systems theory developed in the
previous chapter to a fundamental problem in nonequilibrium
statistical physics, which is to understand the microscopic origin of
diffusion in many-particle systems. We start with a reminder of
diffusion as a simple random walk on the line. Modeling such processes
by suitably generalizing the piecewise linear map studied previously,
we will see how diffusion can be generated by microscopic
deterministic chaos. The main result will be an exact formula relating
the diffusion coefficient, which characterizes macroscopic diffusion
of particles, to the dynamical systems quantities introduced before.

In Section \ref{sec:detdif}, which draws upon Section~2.1 of
\cite{Kla06}, we explain the basic idea of deterministic diffusion and
introduce our model. Section~\ref{sec:erfdd}, which is partially based
on \cite{RKD,RKdiss,KlDo99}, outlines a method of how to exactly
calculate the diffusion coefficient for such types of dynamical
systems.

\section{What is deterministic diffusion?}\label{sec:detdif}

\begin{figure}[t]
\centerline{\includegraphics[width=9cm,angle=-90]{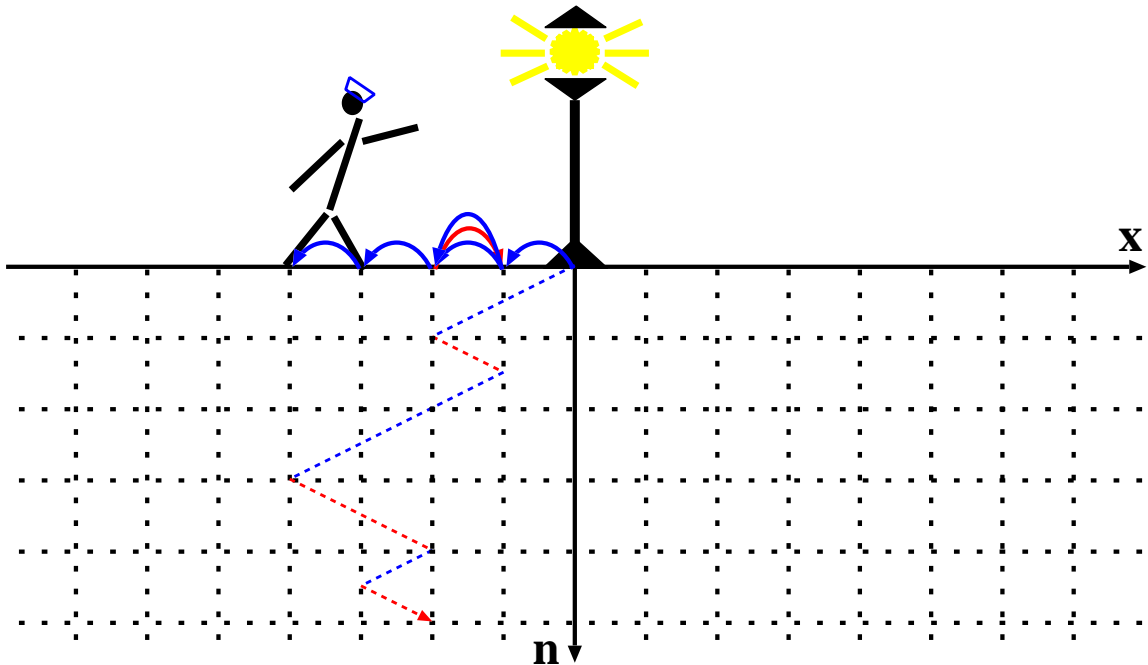}}
\caption{\label{fig:rwdif}The ``problem of the random walk'' in 
  terms of a drunken sailor at a lamppost. The space-time diagram
  shows an example of a trajectory for such a drunken sailor, where
  $n\in\mathbb{N}$ holds for discrete time and $x\in\mathbb{R}$ for
  the position of the sailor on a discrete lattice of spacing $s$. }
\end{figure}

In order to learn about deterministic diffusion, we must first
understand what ordinary diffusion is all about. Here we introduce
this concept by means of a famous example, see Fig.~\ref{fig:rwdif}:
Let us imagine that some evening a sailor wants to walk home, however,
he is completely drunk such that he has no control over his single
steps.  For sake of simplicity let us imagine that he moves in one
dimension.  He starts at a lamppost at position $x=0$ and then makes
steps of a certain step length $s$ to the left and to the right. Since
he is completely drunk he looses all memory between any single steps,
that is, all steps are {\em uncorrelated}.  It is like tossing a coin
in order to decide whether to go to the left or to the right at the
next step. We may now ask for the probability to find the sailor after
$n$ steps at position $x$, i.e., a distance $|x|$ away from his
starting point.

\begin{figure}[t]
\centerline{\includegraphics[width=8cm,angle=-90]{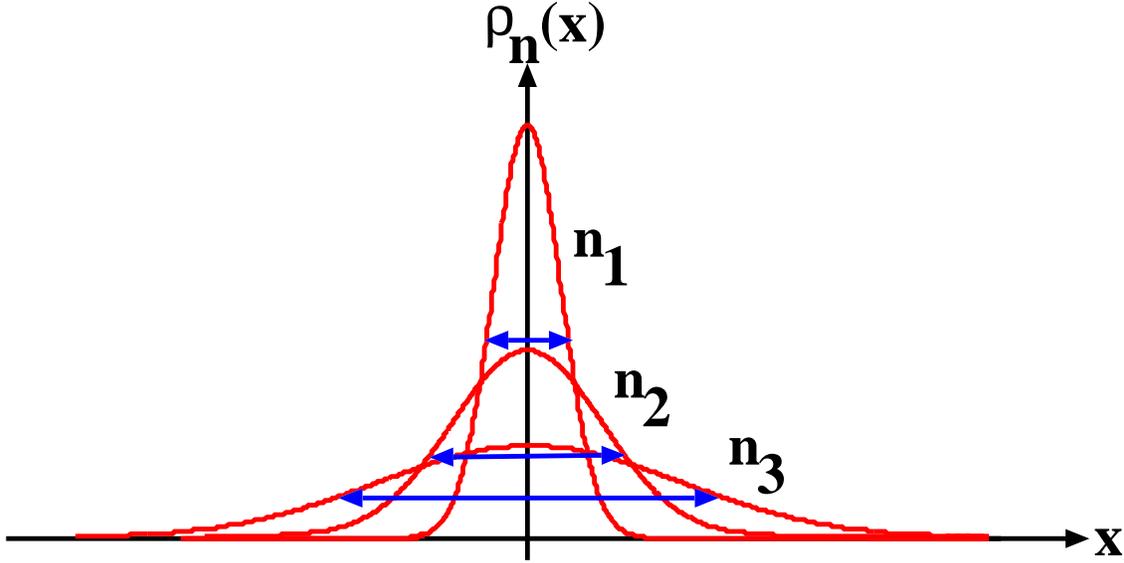}}
\caption{\label{fig:rwpd}Probability distribution functions $\rho_n(x)$ 
  to find a sailor after $n$ time steps at position $x$ on the line,
  calculated for an ensemble of sailors starting at the lamppost,
  cf.\ Fig.~\ref{fig:rwdif}.  Shown are three probability densities
  after different numbers of iteration $n_1<n_2<n_3$.}
\end{figure}

Let us add a short historial note: This ``problem of considerable
interest'' was first formulated (in three dimensions) by Karl Pearson
in a letter to {\em Nature} in 1905 \cite{Pea05}. He asked for a
solution, which was provided by Lord Rayleigh referring to older work
by himself \cite{Ray05}.  Pearson concluded: ``The lesson of Lord
Rayleigh's solution is that in open country the most probable place to
find a drunken man, who is at all capable of keeping on his feet, is
somewhere near his starting point'' \cite{Pea05}. This refers to the
Gaussian probability distributions for the sailor's positions, which
are obtained in a suitable scaling limit from a Gedankenexperiment
with an ensemble of sailors starting from the lamppost.
Fig.~\ref{fig:rwpd} sketches the spreading of such a diffusing
distribution of sailors in time. The mathematical reason for the
emerging Gaussianity of the probability distributions is nothing else
than the central limit theorem \cite{Reif}.

We may now wish to quantify the speed by which a ``droplet of
sailors'' starting at the lamppost spreads out. This can be done by
calculating the {\em diffusion coefficient} for this system. In case
of one-dimensional dynamics the diffusion coefficient can be defined
by the Einstein formula 
\begin{equation}
D:=\lim_{n\to\infty}\frac{1}{2n}<x^2> \quad , \label{eq:rwdk}
\end{equation}
where 
\be
<x^2>:=\int dx\;x^2\rho_n(x) \label{eq:vari}
\ee
is the variance, or second moment, of the probability distribution
$\rho_n(x)$ at time step $n$, also called {\em mean square
displacement} of the particles. This formula may be understood as
follows: For our ensemble of sailors we may choose
$\rho_0(x)=\delta(x)$ as the initial probability distribution with
$\delta(x)$ denoting the (Dirac) $\delta$-function, which mimicks the
situation that all sailors start at the same lamppost at $x=0$. If our
system is ergodic the diffusion coefficient should be independent of
the choice of the initial ensemble. The spreading of the distribution
of sailors is then quantified by the growth of the mean square
displacement in time. If this quantity grows linearly in time, which
may not necessarily be the case but holds true if our probability
distributions for the positions are Gaussian in the long-time limit
\cite{Kla06}, the magnitude of the diffusion coefficient $D$ tells us
how quickly our ensemble of sailors disperses. For further details
about a statistical physics description of diffusion we refer to the
literature \cite{Reif}.

In contrast to this well-known picture of diffusion as a stochastic
random walk, the theory of dynamical systems makes it possible to
treat diffusion as a {\em deterministic dynamical process}.  Let us
replace the sailor by a point particle. Instead of coin tossing, the
orbit of such a particle starting at initial condition $x_0$ may then
be generated by a chaotic dynamical system of the type as considered
in the previous chapters, $x_{n+1}=F(x_n)$.  Note that defining the
one-dimensional map $F(x)$ together with this equation yields the {\em
full microscopic equations of motion} of the system. You may think of
these equations as a caricature of Newton's equations of motion
modeling the diffusion of a single particle. Most importantly, in
contrast to the drunken sailor with his memory loss after any time
step here the {\em complete memory} of a particle is taken into
account, that is, all steps are fully correlated.  The decisive new
fact that distinguishes this dynamical process from the one of a
simple uncorrelated random walk is hence that $x_{n+1}$ is uniquely
determined by $x_n$, rather than having a random distribution of
$x_{n+1}$ for a given $x_n$. If the resulting dynamics of an ensemble
of particles for given equations of motion has the property that a
diffusion coefficient $D>0$ Eq.~(\ref{eq:rwdk}) exists, we speak of
(normal)\footnote{See Section \ref{sec:anodif} for another type of
diffusion, where $D$ is either zero or infinite, which is called {\em
anomalous} diffusion.}  {\em deterministic diffusion}
\cite{Do99,Gasp,Kla06,RKdiss,Schu}.

\begin{figure}[t]
\centerline{\includegraphics[width=12cm]{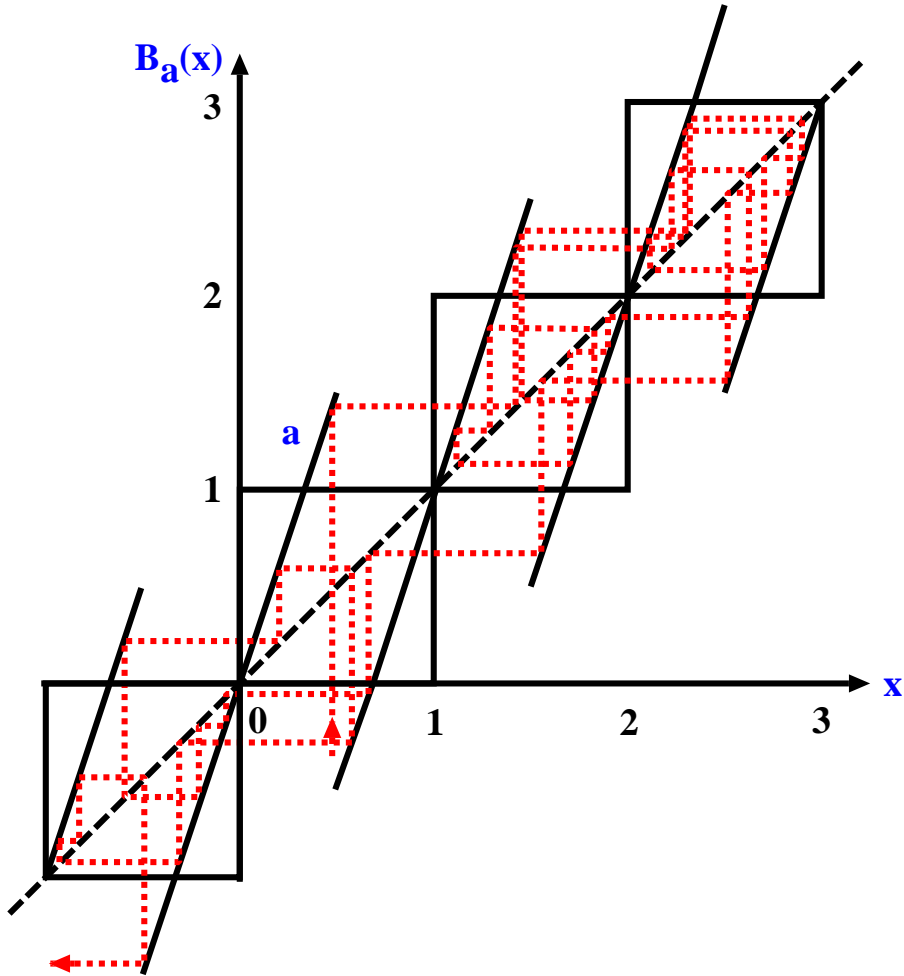}}
\caption{\label{rmbfig1}A simple model for deterministic diffusion. The
  dashed line depicts the orbit of a diffusing particle in form of a
  cobweb plot \cite{ASY97}. The slope $a$ serves as a control
  parameter for the periodically continued piecewise linear map
  $B_a(x)$.}
\end{figure}

Fig.\ \ref{rmbfig1} shows the simple model of deterministic diffusion
that we shall study in this chapter. It depicts a ``chain of boxes''
of chain length $L\in\mathbb{N}$, which continues periodically in both
directions to infinity, and the orbit of a moving point particle. Let
us first specify the map defined on the unit interval, which we may
call the box map. For this we choose the map $B_a(x)$ introduced in
our previous Example~\ref{ex:obern}. We can now periodically continue
this box map onto the whole real line by a {\em lift of degree one},

\begin{equation} 
B_a(x+1)=B_a(x)+1 \quad , \label{eq:lft} 
\end{equation}
for which the acronym {\em old} has been introduced \cite{KaHa95}.
Physically speaking, this means that $B_a(x)$ continued onto the real
line is translational invariant with respect to integers.  Note
furthermore that we have chosen a box map whose graph is point
symmetric with respect to the center of the box at
$(x,y)=(0.5,0.5)$. This implies that the graph of the full map
$B_a(x)$ is anti-symmetric with respect to $x=0$,
\begin{equation}
B_a(x)=-B_a(-x)\quad , \quad \label{eq:sym}
\end{equation}
so that there is no ``drift'' in this chain of boxes. The drift case
with broken symmetry could be studied as well \cite{Kla06}, but we
exclude it here for sake of simplicity.

\section{Escape rate formalism for deterministic diffusion}\label{sec:erfdd}

Before we can start with this method, let us remind ourselves about
the elementary theory of diffusion in form of the diffusion
equation. We then outline the basic idea underlying the escape rate
formalism, which eventually yields a simple formula expressing
diffusion in terms of dynamical systems quantities. Finally, we work
out this approach for our deterministic model introduced before.

\subsection{The diffusion equation}\label{sec:diffeq}

In the last section we have sketched in a nutshell what, in our
setting, we mean if we speak of diffusion. This picture is made more
precise by deriving an equation that exactly generates the dynamics of
the probability densities displayed in Fig.~\ref{fig:rwpd} \cite{Reif}.
For this purpose, let us reconsider for a moment the situation
depicted in Fig.~\ref{fig:phep}. There, we had a gas with an initially
very high concentration of particles on the left hand side of the box.
After the piston was removed, it seemed natural that the particles
spread out over the right hand side of the box as well thus
diffusively covering the whole box. We may thus come to the conclusion
that, firstly, {\em there will be diffusion if the density of
particles in a substance is non-uniform in space}. For this density of
particles and by restricting ourselves to diffusion in one dimension
in the following, let us write $\tilde{n}=\tilde{n}(x,t)$, which holds
for the number of particles that we can find in a small line element
$dx$ around the position $x$ at time step $t$ divided by the total
number of particles $N$.\footnote{Note the fine distinction between
$\tilde{n}(x,t)$ and our previous $\rho_n(x)$, Eq.(\ref{eq:pdfdef}),
in that here we consider continuous time $t$ for the moment, and all
our particles may interact with each other.}

As a second observation, we see that {\em diffusion occurs in the
direction of decreasing particle density}. This may be expressed as
\be
j=:-D\pard{\tilde{n}}{x}\quad ,
\ee
which according to Einstein's formula Eq.~(\ref{eq:rwdk})
may be considered as a second definition of the diffusion coefficient
$D$. Here the flux $j=j(x,t)$ denotes the number of particles passing
through an area perpendicular to the direction  of diffusion per time $t$. 
This equation is known as {\em Fick's first law}.
Finally, let us assume that no particles are created or destroyed
during our diffusion process. In other words, we have {\em
conservation of the number of particles} in form of 
\be
\pard{\tilde{n}}{t}+\pard{j}{x}=0 \quad .
\ee
This {\em continuity equation} expresses the fact that whenever the particle
density $\tilde{n}$ changes in time $t$, it must be due to a spatial
change in the particle flux $j$. Combining the equation with Fick's
first law we obtain {\em Fick's second law},
\be
\pard{\tilde{n}}{t}=D\pard{^2\tilde{n}}{x^2}\quad , \label{eq:diffeq}
\ee
which is also known as the {\em diffusion equation}. Mathematicians
call the process defined by this equation a {\em Wiener process},
whereas physicists rather speak of {\em Brownian motion}. If we would
now solve the diffusion equation for the drunken sailor initial
density $\tilde{n}(x,0)=\delta(x)$, we would obtain the precise
functional form of our spreading Gaussians in Fig.~\ref{fig:rwpd},
\be
\tilde{n}(x,t)=\frac{1}{\sqrt{4\pi Dt}}\exp{\left(-\frac{x^2}{4Dt}\right)} \quad .
\ee
Calculating the second moment of this distribution according to
Eq.~(\ref{eq:vari}) would lead us to recover Einstein's definition of
the diffusion coefficient Eq.~(\ref{eq:rwdk}). Therefore, both this
definition and the one provided by Fick's first law are consistent
with each other.

\subsection{Basic idea of the escape rate formalism}

We are now fully prepared for establishing an interesting link between
dynamical systems theory and statistical mechanics. We start with a
brief outline of the concept of this theory, which is called the {\em
escape rate formalism}, pioneered by Gaspard and others
\cite{GN,Gasp,Do99}. It consists of three steps:

{\bf Step 1:} {\em Solve the one-dimensional diffusion equation}
Eq.~(\ref{eq:diffeq}) derived above {\em for absorbing boundary
conditions}. That is, we consider now some type of {\em open} system
similar to what we have studied in the previous chapter. We may thus
expect that the total number of particles $N(t):=\int dx\:\tilde{n}(x,t)$
within the system decreases exponentially as time evolves according to
the law expressed by Eq.~(\ref{eq:partdec}), that is, 
\be
N(t)=N(0) e^{-\gamma_{de} t}\quad . 
\ee
It will turn out that the escape rate
$\gamma_{de}$ defined by the diffusion equation with absorbing
boundaries is a function of the system size $L$ and of the diffusion
coefficient $D$.

{\bf Step 2:} {\em Solve the Frobenius-Perron equation} 
\be
\rho_{n+1}(x)=\int dy \: \rho_n(y) \:\delta(x-F(y)) \quad ,
\label{eq:fppur} 
\ee 
which represents the continuity equation for the probability density
$\rho_n(x)$ of the map $F(x)$ \cite{Ott,Beck,Do99}, for the very same
{\em absorbing boundary conditions} as in Step~1. Let us assume that
the dynamical system under consideration is normal diffusive, that is,
that a diffusion coefficient $D>0$ exists.  We may then expect a
decrease in the number of particles that is completely analogous to
what we have obtained from the diffusion equation. That is, if we
define as before $N_n:=\int dx\:\rho_n(x)$ as the total number of
particles within the system at discrete time step $n$, in case of
normal diffusion we should obtain
\be
N_n=N_0 e^{-\gamma_{FP} n} \quad . 
\ee
However, in contrast to Step~1 here the escape rate $\gamma_{FP}$
should be fully determined by the dynamical system that we are
considering. In fact, we have already seen before that for open
systems the escape rate can be expressed exactly as the difference
between the positive Ljapunov exponent and the KS-entropy on the
fractal repeller, cf.\ the escape rate formula Eq.~(\ref{eq:escrf}).

{\bf Step 3:} If the functional forms of the particle
density $\tilde{n}(x,t)$ of the diffusion equation and of the probability density
$\rho_n(x)$ of the map's Frobenius-Perron equation {\em match in the limit
of system size and time going to infinity} --- which is what one has 
to show ---, the escape rates
$\gamma_{de}$ obtained from the diffusion equation and $\gamma_{FP}$
calculated from the Frobenius-Perron equation should be equal,
\be
\gamma_{de}=\gamma_{FP} \quad ,
\ee 
providing a fundamental link between the statistical physical theory
of diffusion and dynamical systems theory. Since $\gamma_{de}$ is a
function of the diffusion coefficient $D$, and knowing that
$\gamma_{FP}$ is a function of dynamical systems quantities, we should
then be able to express $D$ exactly in terms of these dynamical
systems quantifiers. We will now illustrate how this method works by
applying it to our simple deterministic diffusive model
introduced above.

\subsection{The escape rate formalism worked out for a simple map}

\begin{figure}[t]
\centerline{\includegraphics[width=9cm]{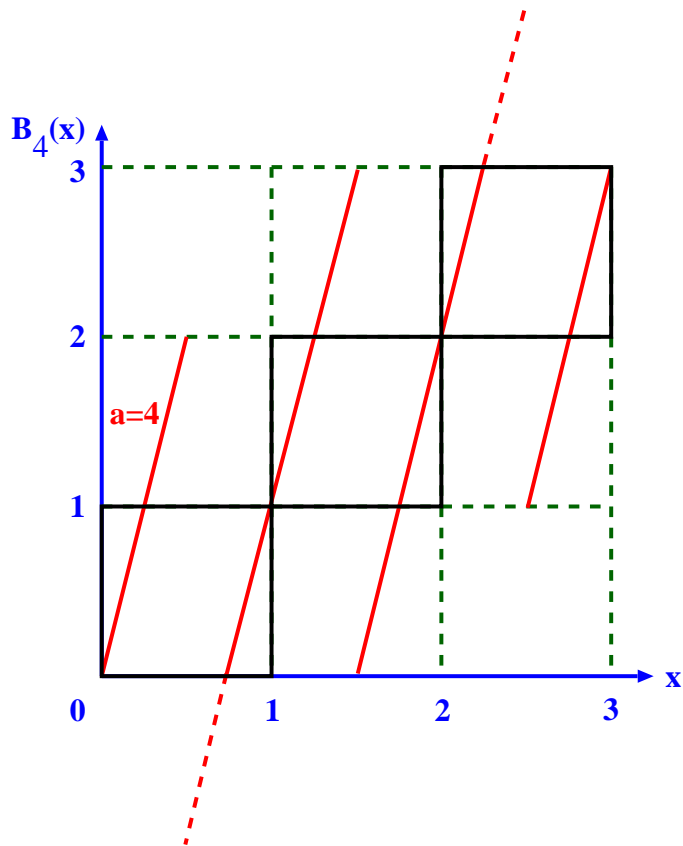}} 
\caption{Our
previous map $B_a(x)$ periodically continued onto the whole real line
for the specific parameter value $a=4$. The example shown depicts a
chain of length $L=3$. The dashed quadratic grid indicates a Markov
partition for this map.}  \label{fig:map4}
\end{figure}

Let us consider the map $B_a(x)$ lifted onto the whole real line for
the specific parameter value $a=4$, see Fig.~\ref{fig:map4}. With $L$
we denote the chain length. Proceeding along the above lines, let us
start with

{\bf Step 1:} Solve the one-dimensional diffusion equation
Eq.~(\ref{eq:diffeq}) for the {\em absorbing boundary conditions}
\be
\tilde{n}(0,t)=\tilde{n}(L,t)=0 \quad , 
\ee
which models the situation that particles escape precisely at the
boundaries of our one-dimensional domain. A straightforward
calculation yields
\be
\tilde{n}(x,t)=\sum_{m=1}^{\infty}b_m\exp\left(-\left(\frac{m\pi}{L}\right)^2Dt\right)
\sin\left(\frac{m\pi}{L}x\right) \label{eq:desoln}
\ee
with $b_m$ denoting the Fourier coefficients.

{\bf Step 2:} Solve the Frobenius-Perron equation Eq.~(\ref{eq:fppur}) for the same
absorbing boundary conditions,
\be
\rho_n(0)=\rho_n(L)=0\quad .
\ee
In order to do so, we first need to introduce a concept called {\em
Markov partition} for our map $B_4(x)$:

\begin{defi} Markov partition, verbally \cite{Beck,dslnotes}

For one-dimensional maps acting on compact intervals a partition is called
{\em Markov} if {\em parts of the partition} get mapped again onto
{\em parts of the partition}, or onto {\em unions of parts of the
partition}.  

\end{defi}

\begin{example}

The dashed quadratic grid in Fig.~\ref{fig:map4} defines a Markov
partition for the lifted map $B_4(x)$.

\end{example}

Of course there also exists a precise formal definition of Markov
partitions, however, here we do not elaborate on these technical
details \cite{dslnotes}.  Having a Markov partition at hand enables us
to rewrite the Frobenius-Perron equation in form of a matrix equation,
where a Frobenius-Perron matrix operator acts onto probability density
vectors defined with respect to this special
partitioning.\footnote{Implicitly we herewith choose a specific space
of functions, which are tailored for studying the statistical dynamics
of our piecewise linear maps; see \cite{JePo04} for mathematical
details on this type of methods.} In order to see this, consider an
initial density of points that covers, e.g., the interval in the
second box of Fig.~\ref{fig:map4} uniformly. By applying the map onto
this density, one observes that points of this interval get mapped
two-fold onto the interval in the second box again, but that there is
also escape from this box which uniformly covers the third and the
first box intervals, respectively. This mechanism applies to any box
in our chain of boxes, modified only by the absorbing boundary
conditions at the ends of the chain of length $L$. Taking into account
the stretching of the density by the slope $a=4$ at each iteration,
this suggests that the Frobenius-Perron equation Eq.~(\ref{eq:fppur})
can be rewritten as
\be
\mbox{\boldmath $\rho$}_{n+1}=\frac{1}{4} \, T(4) \, \mbox{\boldmath
$\rho$}_n \quad , \label{eq:fpm}
\ee
where the $L$ x $L$-transition matrix $T(4)$ must read
\be
T(4)=
\left( \begin{array}{cccccccc} 2 & 1 & 0 & 0 & \cdots &0 & 0 & 0 \\
1 & 2 & 1 & 0 & 0 & \cdots & 0 & 0 \\
0 & 1 & 2 & 1 & 0 & 0 & \cdots & 0 \\
\vdots & & & \vdots & \vdots & & & \vdots \\
0 & \cdots & 0 & 0 & 1 & 2 & 1 & 0 \\
0 & 0 & \cdots & 0 & 0 & 1 & 2 & 1 \\
0 & 0 & 0 & \cdots & 0 & 0 & 1 & 2 \\ 
\end{array} \right) \quad . \label{eq:tm4}
\ee
Note that in any row and in any column we have three non-zero matrix
elements except in the very first and the very last rows and columns,
which reflect the absorbing boundary conditions.  In Eq.(\ref{eq:fpm})
this transition matrix $T(4)$ is applied to a column vector
$\mbox{\boldmath $\rho$}_n$ corresponding to the probability density
$\rho_n(x)$, which can be written as
\be
\mbox{\boldmath $\rho$}_n=|\rho_n(x)>:=(\rho_n^1,\rho_n^2,\ldots,
\rho_n^k,\ldots,\rho_n^L)^*\quad , \label{eq:pdvec}
\ee
where ``$*$'' denotes the transpose and $\rho_n^k$ represents the
component of the probability density in the $k$th box,
$\rho_n(x)=\rho_n^k\;,\; k-1< x\le k\;,\;k=1,\ldots,L\;$, $\rho_n^k$
being constant on each part of the partition.
We see that this transition matrix is symmetric, hence it
can be diagonalized by spectral decomposition. Solving the
eigenvalue problem
\be
T(4)\,|\phi_m(x)>=\chi_m(4)\,|\phi_m(x)>\quad , \label{eq:ewpt4}
\ee
where $\chi_m(4)$ and $|\phi_m(x)>$ are the eigenvalues and
eigenvectors of $T(4)$, respectively, one obtains
\bna
|\rho_n(x)>&=&\frac{1}{4}\sum_{m=1}^{L}\chi_m(4)\,
|\phi_m(x)><\phi_m(x)|\rho_{n-1}(x)> \nonumber \\
&=& \sum_{m=1}^{L}\exp\left(-n\ln\frac{4}{\chi_m(4)}\right) \,
|\phi_m(x)><\phi_m(x)|\rho_0(x)> \quad , \label{eq:pd4}
\ena
where $|\rho_0(x)>$ is the initial probability density vector. Note
that the choice of initial probability densities is restricted by this
method to functions that can be written in the vector form of
Eq.(\ref{eq:pdvec}). It remains to solve the eigenvalue problem
Eq.~(\ref{eq:ewpt4}) \cite{RKD,KlDo99}. The eigenvalue equation for the
single components of the matrix $T(4)$ reads
\begin{equation}
\phi_m^k+2\phi_m^{k+1}+\phi_m^{k+2}=\chi_m \phi_m^{k+1} \quad , \quad
0\le k\le L-1 \quad , \label{eq:phabs}
\end{equation}
supplemented by the absorbing boundary conditions
\be
\phi_m^0=\phi_m^{L+1}=0 \quad . 
\ee
This equation is of the form of a
discretized ordinary differential equation of degree two, hence we make the
ansatz
\begin{equation}
\phi_m^k=a \cos (k\theta) + b \sin (k\theta) \quad , \quad 0\le k \le
L+1 \quad . \label{eq:tmans}
\end{equation}
The two boundary conditions lead to
\begin{equation}
a=0 \quad \mbox{and} \quad \sin ((L+1)\theta)=0
\end{equation}
yielding 
\begin{equation} 
\theta_m=\frac{m \pi}{L+1} \quad , \quad 1\le m\le L \quad .
\end{equation}
The eigenvectors are then determined by
\begin{equation}
\phi_m^k=b \sin(k\theta_m) \quad  . \label{eq:ewp4}
\end{equation}
Combining this equation with Eq.~(\ref{eq:phabs}) yields as the eigenvalues
\begin{equation}
\chi_m=2+2\cos \theta_m \quad . \label{eq:evta}
\end{equation}

{\bf Step 3:} Putting all details together, it remains to match the
solution of the diffusion equation to the one of the Frobenius-Perron
equation: In the limit of time $t$ and system size $L$ to infinity,
the density $\tilde{n}(x,t)$ Eq.~(\ref{eq:desoln}) of the diffusion
equation reduces to the largest eigenmode,
\be
\tilde{n}(x,t)\simeq \exp\left(-\gamma_{de} t\right)B\sin\left(\frac{\pi}{L}x\right)\quad , \label{eq:f2sa}
\ee
where
\be
\gamma_{de} :=\left(\frac{\pi}{L}\right)^2D \label{eq:gammade}
\ee
defines the escape rate  as determined by the diffusion equation.
Analogously, for discrete time $n$ and chain length $L$ to infinity
we obtain for the probability density of the
Frobenius-Perron equation, Eq.(\ref{eq:pd4}) with Eq.(\ref{eq:ewp4}), 
\bna
\rho_n(x)&\simeq& \exp\left(-\gamma_{FP}n\right)\tl{B}\sin\left(\frac{\pi}{L+1}k\right)\quad , \nonumber \\
& & k=0,\ldots,L+1 \quad , \quad k-1< x \le k \label{eq:pd4a}
\ena
with an escape rate of this dynamical system given by
\be
\gamma_{FP}=\ln\frac{4}{2+2\cos(\pi/(L+1))} \quad , \label{eq:escr4}
\ee
which is determined by the largest eigenvalue $\chi_1$ of the matrix
$T(4)$, see Eq.(\ref{eq:pd4}) with Eq.(\ref{eq:evta}). We can now see
that the functional forms of the eigenmodes of Eqs.(\ref{eq:f2sa}) and
(\ref{eq:pd4a}) match precisely.\footnote{We remark that there are
discretization effects in the time and position variables, which are
due to the fact that we compare a time-discrete system defined on a
specific partition with time-continuous dynamics. They disappear in
the limit of time to infinity by using a suitable spatial coarse
graining.}  This allows us to match Eqs.~(\ref{eq:gammade}) and
(\ref{eq:escr4}) leading to
\be
D(4)=\left(\frac{L}{\pi}\right)^2\gamma_{FP} \quad . \label{eq:d4escr}
\ee
Using the right hand side of Eq.~(\ref{eq:escr4}) and expanding it for $L\to\infty$,
this formula enables us to calculate the diffusion coefficient $D(4)$ to
\be
D(4)=\left(\frac{L}{\pi}\right)^2\gamma_{FP} =\frac{1}{4}\frac{L^2}{(L+1)^2}+{\cal O}(L^{-4}) \;
\to \; \frac{1}{4} \quad (L\rightarrow \infty)\quad . \label{eq:dkl}
\ee
Thus we have developed a method by which we can exactly calculate the
deterministic diffusion coefficient of a simple chaotic dynamical
system. However, more importantly, instead of using the explicit
expression for $\gamma_{FP}$ given by Eq.~(\ref{eq:escr4}), let us
remind ourselves of the escape rate formula Eq.~(\ref{eq:escrf}) for
$\gamma_{FP}$,
\be
\gamma_{FP}=\gamma({\cal C}_{B_4})=\lambda({\cal C}_{B_4})-h_{KS}({\cal C}_{B_4}) \quad ,
\ee
which more geneally expresses this escape rate in terms of 
dynamical systems quantities.  Combining this equation with the above
equation Eq.~(\ref{eq:d4escr}) leads to our final result, the {\em
escape rate formula for deterministic diffusion} \cite{GN,Gasp}
\be
D(4)=\lim_{L\to\infty}\left(\frac{L}{\pi}\right)^2 \left[\lambda({\cal C}_{B_4})-h_{KS}({\cal C}_{B_4})\right] \quad .
\ee
We have thus established a fundamental link between quantities
assessing the chaotic properties of dynamical systems and the
statistical physical property of diffusion.

\begin{remark} 
  
\nl
  
\vspace*{-0.7cm}

\begin{enumerate}

\item Above we have only considered the special case of the control
parameter $a=4$. Along the same lines, the diffusion coefficient can
be calculated for other parameter values of the map
$B_a(x)$. Surprisingly, the parameter-dependent diffusion coefficient
of this map turns out to be a fractal function of the control
parameter \cite{RKD,RKdiss,KlDo99}. This result is believed to hold for a
wide class of dynamical systems \cite{Kla06}.

\item The escape rate formula for diffusion does not only hold for
simple one-dimensional maps but can be generalized to
higher-dimensional time-discrete as well as time-continuous dynamical
systems \cite{Gasp,Do99}.

\item This approach can also be generalized to establish relations
between chaos and transport for other transport coefficients such as
viscosity, heat conduction and chemical reaction rates \cite{Gasp}.

\item This is not the only approach connecting transport properties
with dynamical systems quantities. In recent research it was found
that there exist at least two other ways to establish relations that
are different but of a very similar nature; see \cite{Kla06} for
further details.

\end{enumerate}
\end{remark}

\chapter{Anomalous diffusion}\label{chap:anodif}

In the last section we have explored diffusion for a simple piecewise
linear map. One may now wonder what type of diffusive behavior is
encountered if we consider more complicated models. Straightforward
generalizations are nonlinear maps generating {\em intermittency}. In
Section \ref{sec:anodif} we briefly illustrate the phenomenon of
intermittency and introduce the concept of {\em anomalous
diffusion}. We then give an outline of {\em continuous time random
walk theory}, which is a powerful tool of stochastic theory that
models anomalous diffusion. By using this method we derive a {\em
fractional diffusion equation}, which generalizes Fick's second law
that we have encountered before to this type of anomalous diffusion.

After having restricted ourselves to rather abstract but mostly
solvable models, we conclude our review by discussing an experiment
which gives evidence for the existence of anomalous diffusion in a
fundamental biological process. Section \ref{sec:anocell} first
motivates the problem of cell migration. We then present experimental
results for two different cell types and explain them by suggesting a
model reproducing the observed anomalous dynamics of cell migration.

Section \ref{sec:anodif} particularly draws on
Refs.~\cite{KCKSG06,KKCSG07}, see also Section 6.2 of
\cite{Kla06}, Section \ref{sec:anocell} is based on 
Ref.~\cite{DKPS08}. For more general introductions to the very active
field of anomalous transport see, e.g.,
Refs.~\cite{SZK93,KSZ96,MeKl00,KRS08}.

\section{Anomalous diffusion in intermittent maps}\label{sec:anodif}

\subsection{What is anomalous diffusion?}

\begin{figure}[t]
\centerline{\includegraphics[height=7cm]{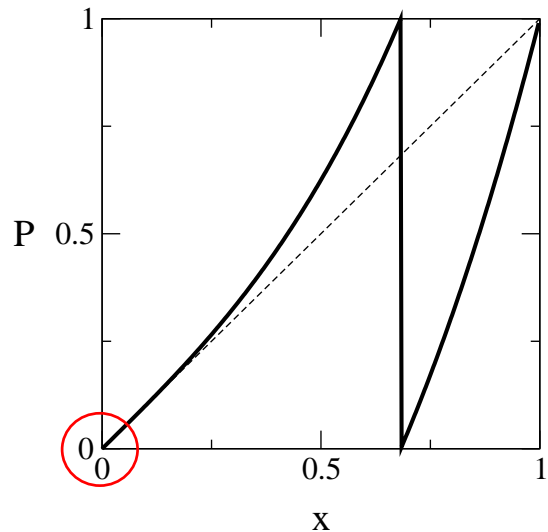}}
\caption{The Pomeau-Manneville map Eq.~(\ref{eq:pmmap}) for $a=1$ 
  and $z=3$.  Note that there is a marginal fixed point at $x=0$
  leading to the intermittent behavior depicted in
  Fig.~\ref{fig:ipoma}.}
\label{fig:poma} 
\end{figure}

\begin{figure}[t]
  \centerline{\includegraphics[width=15cm]{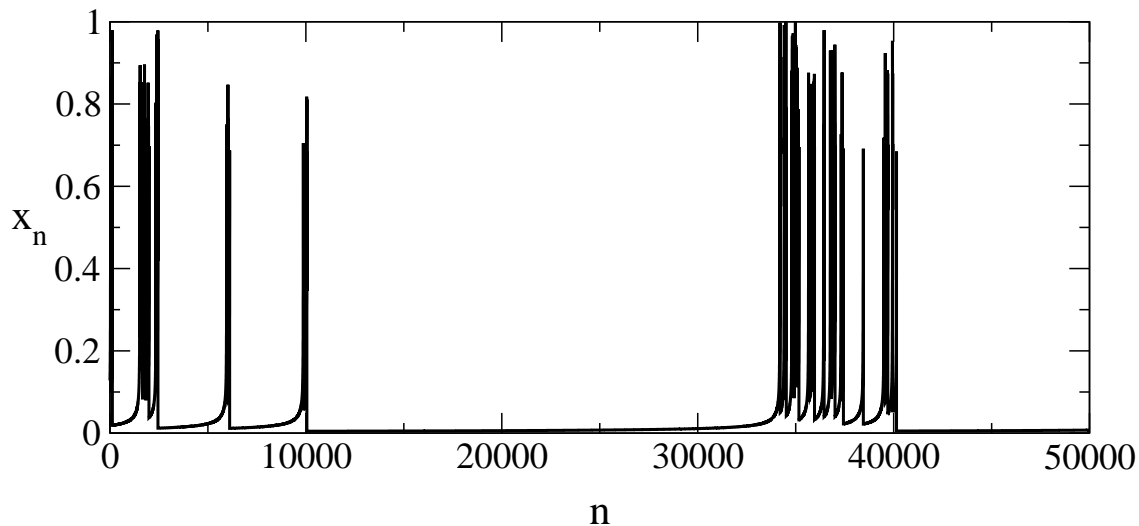}}
\caption{Phenomenology of intermittency in the Pomeau-Manneville map
  Fig.~\ref{fig:poma}: The plot shows the time series of position
  $x_n$ versus discrete time step $n$ for an orbit generated by the
  map Eq.~(\ref{eq:pmmap}), which starts at a typical initial condition
  $x_0$.}
\label{fig:ipoma} 
\end{figure}

Let us consider a simple variant of our previous piecewise linear
model, which is the {\em Pomeau-Manneville map} \cite{PoMa80}
\be
P_{a,z}(x) = x + ax^z \; \; \mbox{mod} 1\quad , \label{eq:pmmap}
\ee
see Fig.~\ref{fig:poma}, where as usual the dynamics is defined by
$x_{n+1} = P_{a,z}(x_n)$. This map has two control parameters, $a\ge
1$ and the exponent of nonlinearity $z\ge 1$.  For $a=1$ and $z=1$
this map just reduces to our familiar Bernoulli shift
Eq.~(\ref{eq:bern}), however, for $z>1$ it provides a nontrivial
nonlinear generalization of it. The nontriviality is due to the fact
that in this case the stability of the fixed point at $x=0$ becomes
{\em marginal} (sometimes also called indifferent, or neutral),
$P'_{a,z}(0)=1$. Since the map is smooth around $x=0$, the dynamics
resulting from the left branch of the map is determined by the
stability of this fixed point, whereas the right branch is just of
Bernoulli shift-type yielding ordinary chaotic dynamics. There is thus
a competition in the dynamics between these two different branches as
illustrated by Fig.~\ref{fig:ipoma}: One can observe that long
periodic {\em laminar phases} determined by the marginal fixed point
around $x=0$ are interrupted by {\em chaotic bursts} reflecting the
``Bernoulli shift-like part'' of the map with slope $a>1$ around
$x=1$. This phenomenology is the hallmark of what is called {\em
intermittency} \cite{Schu,Ott}.

\begin{figure}[t]
\centerline{\includegraphics[width=9.7cm]{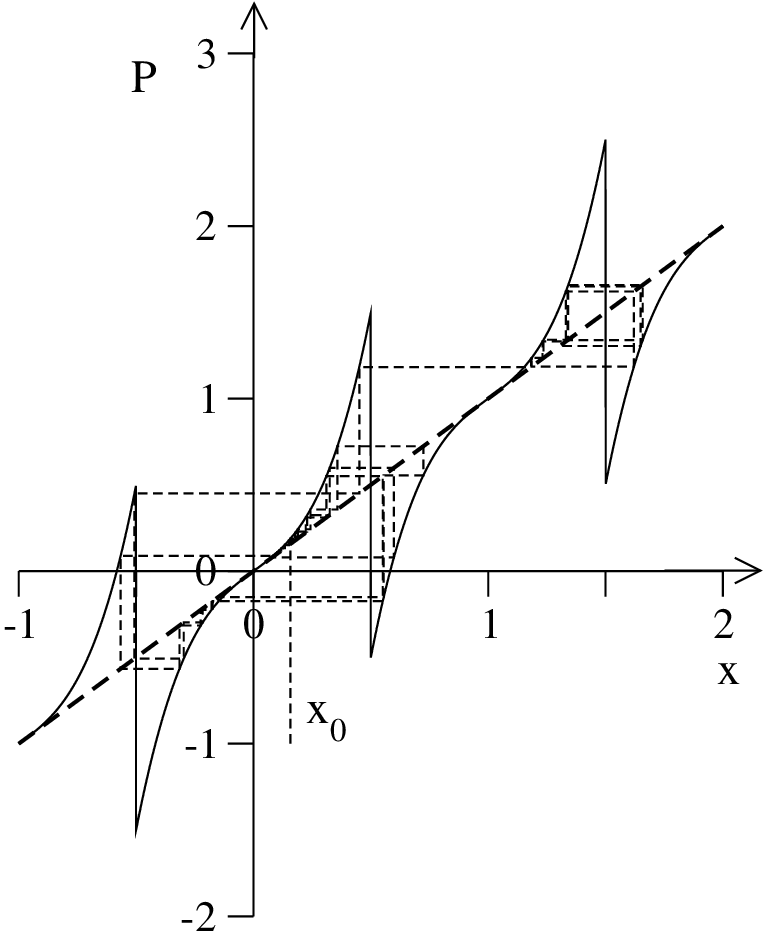}}
\caption{The Pomeau-Manneville map Fig.~\ref{fig:poma}, Eq.~(\ref{eq:pmmap}), 
  lifted symmetrically onto the whole real line such that it generates
  subdiffusion.}
\label{fig:liftedpm} 
\end{figure}

Following Section~\ref{sec:detdif} it is now straightforward to define
a spatially extended version of the Pomeau-Manneville map: For this
purpose we just continue $P_{a,z}(x) = x +a x^z\;,\; 0 \le x
<\frac{1}{2}$ in Eq.~(\ref{eq:pmmap}) onto the real line by the
translation $P_{a,z}(x + 1) = P_{a,z} (x) + 1$, see
Eq.~(\ref{eq:lft}), under reflection symmetry
$P_{a,z}(-x)=-P_{a,z}(x)$, see Eq.~(\ref{eq:sym}). The resulting model
~\cite{GeTo84,ZuKl93a} is displayed in Fig.~\ref{fig:liftedpm}. As
before, we may now be interested in the type of deterministic
diffusion generated by this model. Surprisingly, by calculating the
mean square displacement Eq.~(\ref{eq:vari}) either analytically or
from computer simulations one finds that for $z>2$
  
\be
\left< x^2 \right>\sim\:n^{\alpha}\quad , \quad \alpha<1\quad(n\to\infty)\quad . \label{eq:amsd}
\ee
This implies that the diffusion coefficient
$D:=\lim_{n\to\infty}<x^2>/(2n)$ as defined by Eq.~(\ref{eq:rwdk}) is
simply zero, despite the fact that particles can go anywhere on the
real line as shown in Fig.~\ref{fig:liftedpm}. We thus encounter a
novel type of diffusive behavior classified by the following
definition:

\begin{defi} anomalous diffusion \cite{MeKl00,KRS08}
  
  If the exponent $\alpha$ in the temporal spreading of the mean
  square displacement Eq.~(\ref{eq:amsd}) of an ensemble of particles
  is not equal to one, one speaks of {\em anomalous diffusion}. If
  $\alpha<1$ one says that there is {\em subdiffusion}, for $\alpha>1$
  there is {\em superdiffusion}, and in case of $\alpha=1$ one refers
  to {\em normal diffusion}. The constant 
\be
K:=\lim_{n\to\infty}\frac{<x^2>}{n^{\alpha}}\quad , \label{GDC_Def}
\ee
  where in case of normal diffusion $K=2D$, is called the {\em
  generalized diffusion coefficient}.\footnote{In detail, the
  definition of a generalized diffusion coefficient is a bit more
  subtle \cite{KKCSG07}.}

\end{defi}

We will now discuss how $K$ behaves as a function of $a$ for our new
model and then show how the exponent $\alpha$ and, in an
approximation, $K$, can be calculated analytically. This can be
achieved by means of {\em continuous time random walk} (CTRW) theory,
which provides a generalization of the drunken sailor's model
introduced in Section~\ref{sec:detdif} to anomalous dynamics.

\subsection{Continuous time random walk theory}

\begin{figure}[t]
\centerline{\includegraphics[width=11cm]{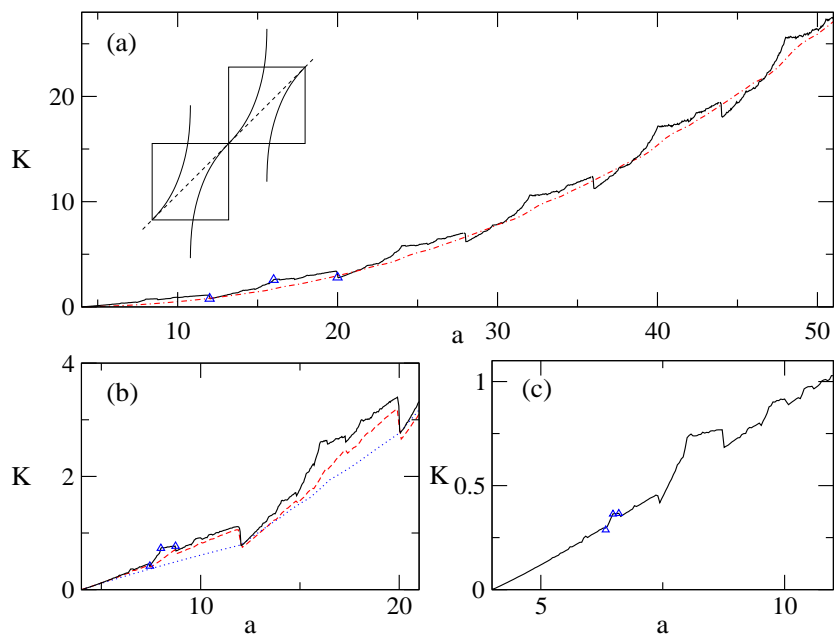}}
\caption{The generalized diffusion coefficient $K$ Eq.~(\ref{GDC_Def})
  as a function of $a$ for $z=3$. The curve in (a) consists of $1200$
  points, the dashed-dotted line displays the CTRW result $K_1$,
  Eqs.~(\ref{jump_length}), (\ref{GDC_CTRW}). (b) (600 points) and (c)
  (200 points) show magnifications of (a) close to the onset of
  diffusion.  The dotted line in (b) is the CTRW approximation $K_2$,
  Eqs.~(\ref{int_jump_length}), (\ref{GDC_CTRW}), the dashed line
  represents yet another semi-analytical approximation as detailed in
  Refs.~\cite{KCKSG06,KKCSG07}. The triangles mark a specific
  structure appearing on finer and finer scales. The inset in (a)
  depicts again the model Eq.~(\ref{eq:pmmap}).}
\label{fig:fctrw2}
\end{figure}

Let us first study the diffusive behavior of the map displayed in
Fig.~\ref{fig:liftedpm} by computer simulations.\footnote{All
simulations were performed starting from a uniform, random
distribution of $10^6$ initial conditions on the unit interval by
iterating for $n=10^4$ time steps.} As we will explain in detail
below, stochastic theory predicts that for this map it is 
\begin{equation}
\alpha = \begin{cases} 1 , & 1 \le z < 2 \cr \label{eq:alpha}
\frac{1}{z-1}, & 2\le z  \end{cases} 
\end{equation} 
\cite{GeTo84,ZuKl93a}. For all values of the second control 
parameter $a$ we indeed find excellent agreement between these
analytical solutions and the results for $\alpha$ obtained from
simulations.  Consequently, $\alpha$ is determined by
Eq.~(\ref{eq:alpha}) in the following for extracting the generalized
diffusion coefficient $K$ Eq.~(\ref{GDC_Def}) from simulations.

While in Chapter~\ref{chap:detdif} we have only discussed diffusion
for a specific choice of the control parameter, we now study the
behavior of $K$ as a function of $a$ for fixed $z$. Computer
simulation results are displayed in
Fig.~\ref{fig:fctrw2}. Magnifications of part (a) shown in parts (b)
and (c) reveal self similar-like irregularities indicating a fractal
parameter dependence of $K=K(a)$. This fractality is highlighted by
the sub-structure identified through triangles, which is repeated on
finer and finer scales. The parameter values for these symbols
correspond to specific series of Markov partitions. Details are
explained in Refs.~\cite{RKD,RKdiss,KlDo99} for parameter-dependent
diffusion in the piecewise linear one-dimensional maps studied in
Chapter~\ref{chap:detdif}, which exhibits quite analogous
structures. Over the past few years such fractal transport
coefficients have been revealed for a number of different models. They
are conjectured to be a typical phenomenon if the dynamical system is
deterministic, low-dimensional and spatially periodic. Their origin
can be understood in terms of microscopic long-range dynamical
correlations that, due to topological instabilities of dynamical
systems, change in a complicated manner under parameter
variation. Although it has been argued that such highly irregular
behavior of transport coefficients should also occur in physically
realistic systems, it has not yet clearly been observed in
experiments; see Ref.~\cite{Kla06} for a review of this line of
research.

Instead of elaborating on the fractality in detail, here we reproduce
the course functional form of $K(a)$ by using stochastic CTRW theory.
Pioneered by Montroll, Weiss and Scher \cite{MW65,MS73,SM75}, it
yields perhaps the most fundamental theoretical approach to explain
anomalous diffusion \cite{BoGe90,Weiss94,EbSo05}. In further
groundbreaking works by Geisel et al.\ and Klafter et al., this method
was then adapted to sub- and superdiffusive deterministic maps
\cite{GeTo84,GNZ85,ShlKl85,ZuKl93a}

The basic assumption of the approach is that diffusion can be
decomposed into two stochastic processes characterized by waiting
times and jumps, respectively. Thus one has two sequences of
independent identically distributed random variables, namely a
sequence of positive random waiting times $T_1, T_2, T_3,
\ldots$ with probability density function $w(t)$ and a sequence of
random jumps $\zeta_1, \zeta_2,
\zeta_3, \ldots$ with a probability density function $\lambda(x)$. For
example, if a particle starts at point $x = 0$ at time $t_0 = 0$ and
makes a jump of length $\zeta_n$ at time $t_n = T_1 + T_2 + ... +
T_n$, its position is $x = 0$ for $0 \le t < T_1 = t_1$ and $x =
\zeta_1 + \zeta_2 + ...  + \zeta_n$ for $t_n \le t < t_{n+1}$. The
probability that at least one jump is performed within the time
interval $[0,t)$ is then $\int_0^t dt' w(t')$ while the probability
for no jump during this time interval reads $\Psi(t) = 1 -
\int_{0}^{t} dt' w(t')$. The master equation for the probability
density function $P(x,t)$ to find a particle at position $x$ and time
$t$ is then
\begin{equation}
  \label{master_eq} P(x,t) = \int_{-\infty}^{\infty} dx' \lambda (x
  - x') \int_{0}^{t} dt' \; w(t -  t') \; P(x',t') + \Psi(t) \delta(x)\quad .
\end{equation}
It has the following probabilistic meaning: The probability density
function to find a particle at position $x$ at time $t$ is equal to
the probability density function to find it at point $x'$ at some
previous time $t'$ multiplied with the transition probability to get
from $(x',t')$ to $(x,t)$ integrated over all possible values of $x'$
and $t'$. The second term accounts for the probability of remaining at
the initial position $x=0$. The most convenient representation of this
equation is obtained in terms of the Fourier-Laplace transform of the
probability density function,
\begin{equation} \label{times_PDF} 
\hat{\tilde{P}} (k,s) =
\int_{-\infty}^{\infty} dx \; e^{i k x} \int_{0}^{\infty} dt \;
e^{-st} P(x,t) \quad , 
\end{equation}
where the hat stands for the Fourier transform and the tilde for the
Laplace transform. Respectively, this function obeys the
Fourier-Laplace transform of Eq.~(\ref{master_eq}), which is called
the Montroll-Weiss equation \cite{MW65,MS73,SM75},
\index{Montroll-Weiss equation|ibold}
\begin{equation} \label{Montroll_Weiss} 
\hat{\tilde{P}} (k,s) =
\frac{1 - \tilde{w}(s)}{s} \frac{1}{1-
\hat{\lambda}(k)\tilde{w}(s)}\quad .  \end{equation}
The Laplace transform of the mean square displacement can be
readily obtained by differentiating the Fourier-Laplace transform
of the probability density function,
\begin{equation}
\label{Laplace_MSD}
\tilde{ \left< x^2 (s) \right> } = \int_{-\infty}^{\infty} dx \; x^2 \tilde{P}(x,s) =
\left. - \frac{\partial^2 \hat{\tilde{P}} (k,s) }{\partial k^2} \right|_{k=0}\quad .
\end{equation}
In order to calculate the mean square displacement within this theory,
it thus suffices to know $\lambda(x)$ and $w(t)$ generating the
stochastic process. For one-dimensional maps of the type of
Eq.~(\ref{eq:pmmap}), exploiting the symmetry of the map the waiting
time distribution can be calculated from the approximation
\begin{equation}
\label{cont_time_map}
x_{n+1} - x_n \simeq \frac{dx_t}{dt} = a x_t^z, \; \; x \ll 1 \quad ,
\end{equation}
where we have introduced the continuous time $t\ge 0$. This equation
can easily be solved for $x_t$ with respect to an initial condition
$x_0$. Now one needs to define when a particle makes a ``jump'', as
will be discussed below. By inverting the solution for $x_t$, one can
then calculate the time $t$ a particle has to wait before it makes a
jump as a function of the initial condition $x_0$. This information
determines the relation between the waiting time probability density
\index{waiting time distribution}
$w(t)$ and the as yet unknown probability density of injection points,
\begin{equation}
\label{cont_time_map_sol4}
w(t) \simeq P_{in}(x_0) \left| \frac{dx_0}{dt}\right| \quad .
\end{equation}
Making the assumption that the probability density of injection points
is uniform, $P_{in} \simeq 1$, the waiting time probability density is
straightforwardly calculated from the knowledge of $t(x_0)$.
The second ingredient that is needed for the CTRW approach is the jump
probability density. Standard CTRW theory takes jumps between
neighbouring cells {\em only} into account leading to the ansatz
\cite{GeTo84,ZuKl93a}
\begin{equation}
\label{jump_PDF}
\lambda (x) = \delta (|x| - 1) \quad .
\end{equation}
It turns out that a correct application of this theory to our results
for $K(a)$ requires a modification of the standard theory at three
points: Firstly, the waiting time probability density function must be
calculated according to the grid of elementary cells indicated in
Fig.~\ref{fig:fctrw2}
\cite{RKdiss,dcrc} yielding
\begin{equation}
 w(t)=a\left( 1+a(z-1)t\right) ^{-\frac z{z-1}}\quad .
\label{waiting_time_pdf}
\end{equation}
However, this probability density function also accounts for {\em
  attempted} jumps to another cell, since after a step the particle
may stay in the same cell with a probability of $(1 - p)$.  The latter
quantity is roughly determined by the size of the escape region $p = (
1 - 2 x_c )$ with $x_c$ as a solution of the equation $x_c + a
x^{z}_{c} = 1$. We thus model this fact, secondly, by a jump length
distribution in the form of
\begin{equation}
\label{real_jump_PDF}
\lambda (x) = \frac{p}{2} \delta (\left| x \right| - l) + (1 - p) \delta
(x)\:.
\end{equation}
Thirdly, we introduce two definitions of a typical jump length
$l_i\:,i\in\{1,2\}$.
\begin{equation}
\label{jump_length}
l_1 = \left\{ | M_{a,z}(x) - x | \right\} \:
\end{equation}
corresponds to the actual mean displacement while
\begin{equation}
\label{int_jump_length}
l_{2} = \left\{ | [M_{a,z}(x)] | \right\} \:
\end{equation}
gives the coarse-grained displacement in units of elementary cells, as
it is often assumed in CTRW approaches. In these definitions
$\left\{\ldots\right\}$ denotes both a time and ensemble average over
particles leaving a box. Working out the modified CTRW approximation
sketched above by taking these three changes into account we obtain
for the generalized diffusion coefficient \index{diffusion!coefficient!generalized}
\begin{equation}
K_i  = p l_i^2\left\{
\begin{array}{l@{\quad,\quad}l}
 a^{\gamma} \sin (\pi \gamma) / \pi \gamma^{1+\gamma} & 0 < \gamma < 1 \\
 a (1-1/\gamma) & 1 \le \gamma < \infty
\end{array}\right. \quad ,\label{GDC_CTRW}
\end{equation}
where $\gamma:=1/(z-1)$, which for $z\ge 2$ is identical with $\alpha$
defined in Eq.~(\ref{eq:alpha}). Fig.~\ref{fig:fctrw2} (a) shows that
$K_1$ well describes the coarse functional form of $K$ for large
$a$. $K_2$ is depicted in Fig.~\ref{fig:fctrw2} (b) by the dotted line
and is asymptotically exact in the limit of very small $a$. Hence, the
generalized diffusion coefficient exhibits a dynamical crossover
between two different coarse grained functional forms for small and
large $a$, respectively. An analogous crossover has been reported
earlier for normal diffusion \cite{RKdiss,dcrc} and was also found in
other models \cite{Kla06}.

Let us finally focus on the generalized diffusion coefficient at
$a=12,20,28,...$, which correspond to integer values of the height
$h=[M_{a,z}(1/2)]$ of the map.  Simulations reproduce, within
numerical accuracy, the results for $K_2$ by indicating that $K$ is
discontinuous at these parameter values. Due to the self similar-like
structure of the generalized diffusion coefficient it was thus
conjectured that the precise $K$ of our model exhibits infinitely many
discontinuities on fine scales as a function of $a$, which is at
variance with the continuity of the CTRW approximation
\cite{KCKSG06,KKCSG07}. This highlights again that CTRW 
theory gives only an approximate solution for the generalized
diffusion coefficient of this model.

\subsection{A fractional diffusion equation}

We now turn to the probability density functions (PDFs) generated by
the map Eq.~(\ref{eq:pmmap}). As we will show now, CTRW theory not
only predicts the power $\gamma$ correctly but also the form of the
coarse grained PDF $P(x,t)$ of displacements. Correspondingly the
anomalous diffusion process generated by our model is not described by
an ordinary diffusion equation but by a generalization of it. Starting
from the Montroll-Weiss equation and making use of the expressions for
the jump and waiting time PDFs Eqs.~(\ref{jump_PDF}),
(\ref{waiting_time_pdf}), we rewrite Eq.~(\ref{Montroll_Weiss}) in the
long-time and -space asymptotic form
\begin{equation} 
s^{\gamma}\hat{\tilde{P}} - s^{\gamma - 1} = - \frac{p l_i^2}{2 c b^{\gamma}} k^2
\hat{\tilde{P}} \label{Montroll-Weiss_2} 
\end{equation} 
with $c = \Gamma (1-\gamma)$ and $b=\gamma /a$.  For the initial
condition $P(x,0)=\delta (x)$ of the PDF we have
$\hat{P}(k,0)=1$. Interestingly, the left hand side of this equation
corresponds to the definition of the {\em Caputo fractional
derivative} of a function $G$,
\begin{equation} \frac{\partial^{\gamma}
G}{\partial t^{\gamma}} := \frac{1}{\Gamma (1-\gamma)} \; \int_0^t
dt^{^{\prime }}(t-t^{^{\prime }})^{-\gamma }\frac{\partial G}{\partial
t^{^{\prime }}} \label{Caputo} \quad ,
\end{equation}
in Laplace space \cite{Podl,Mai97},
\begin{equation}
\int_{0}^{\infty} dt \; e^{-st} \frac{\partial^{\gamma} G}{\partial
t^{\gamma}} = s^{\gamma} \tilde{G} (s) - s^{\gamma - 1} G(0) \quad .
\label{Caputo_Laplace} 
\end{equation}
Thus, fractional derivatives come naturally into play as a suitable
mathematical formalism whenever there are power law memory kernels in
space and/or time generating anomalous dynamics; see, e.g.,
Refs.~\cite{SKB02,MeKl00} for short introductions to fractional
derivatives and Ref.~\cite{Podl} for a detailed exposition. Turning
back now to real space, we thus arrive at the time-fractional
diffusion equation
\begin{equation}
\frac{\partial ^\gamma P(x,t)}{\partial t^\gamma } = D \; \frac{\partial ^2P}{\partial x^2}
\label{Frac_Dif_Eq}
\end{equation}
with $D=K\Gamma (1 + \gamma)/2$, $0<\gamma <1$, which is an example of
a {\em fractional diffusion equation} generating subdiffusion. For
$\gamma=1$ we recover the ordinary diffusion equation. The solution of
Eq.~(\ref{Frac_Dif_Eq}) can be expressed in terms of an M-function of
Wright type \cite{Mai97} and reads
\begin{equation}
P(x,t)=\frac 1{2\sqrt{D}t^{\gamma /2}}M\left( \xi,\frac \gamma
2\right) \quad .
\label{ctrw_sol}
\end{equation}

\begin{figure}[t]
\centerline{\includegraphics[width=9.5cm]{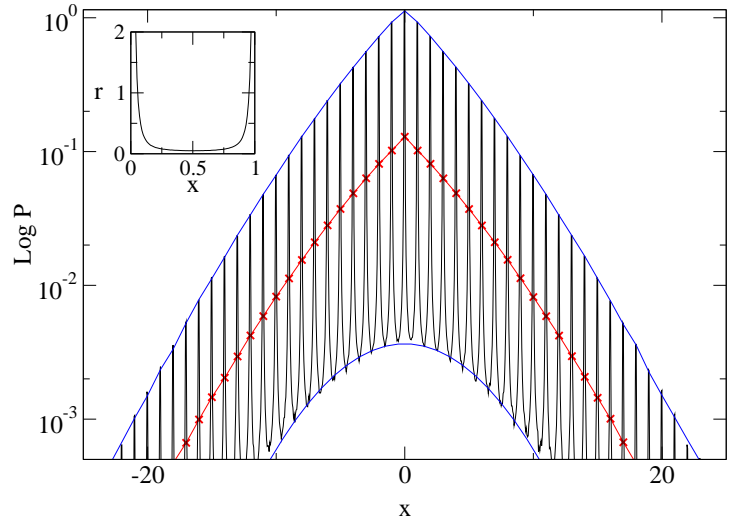}}
\caption{Comparison of the probability density obtained from simulations of 
  the map Eq.~(\ref{eq:pmmap}) (oscillatory structure) with the
  analytical solution Eq.\ (\ref{ctrw_sol}) of the fractional diffusion
  equation Eq.~(\ref{Frac_Dif_Eq}) (continuous line in the middle) for
  $z=3$ and $a=8$. The probability density was computed from $10^7$
  particles after $n=10^3$ iterations. For the generalized diffusion
  coefficient in Eq.~(\ref{ctrw_sol}) the simulation result was used. The
  crosses (x) represent the numerical results coarse grained over unit
  intervals. The upper and the lower curves correspond to fits with a
  stretched exponential and a Gaussian distribution, respectively. The
  inset depicts the probability density function for the map on the
  unit interval with periodic boundaries.}
\label{fig:fctrw4}
\end{figure}

Fig.~\ref{fig:fctrw4} demonstrates an excellent agreement between the
analytical solution Eq.\ (\ref{ctrw_sol}) and the PDF obtained from
simulations of the map Eq.~(\ref{eq:pmmap}) if the PDF is coarse
grained over unit intervals.  However, it also shows that the coarse
graining eliminates a periodic fine structure that is not captured by
Eq.~(\ref{ctrw_sol}). This fine structure derives from the
``microscopic'' PDF of an elementary cell (with periodic boundaries)
as represented in the inset of Fig.~\ref{fig:fctrw4}
\cite{RKdiss}. The singularities are due to the marginal fixed points
of the map, where particles are trapped for long times.  Remarkably,
that way the microscopic origin of the intermittent dynamics is
reflected in the shape of the PDF on the whole real line: From
Fig.~\ref{fig:fctrw4} it is seen that the oscillations in the PDF are
bounded by two functions, the upper curve being of a stretched
exponential type while the lower is Gaussian. These two envelopes
correspond to the laminar and chaotic parts of the motion,
respectively.\footnote{The two envelopes shown in Fig.\
\ref{fig:fctrw4} represent fits with the Gaussian $a_0 \exp (-
x^2/a_1)$ and with the M-function $b_0 \; M(\left| x \right|/ b_1,
\frac{\gamma}{2})$, where $a_0=0.0036$, $a_1=55.0183$ and $b_0 =
3.05$, $b_1 = 0.37$.}

\section{Anomalous diffusion of migrating biological cells} \label{sec:anocell}

\subsection{Cell migration}

\begin{figure}[t]  
\centerline{\includegraphics[height=8cm]{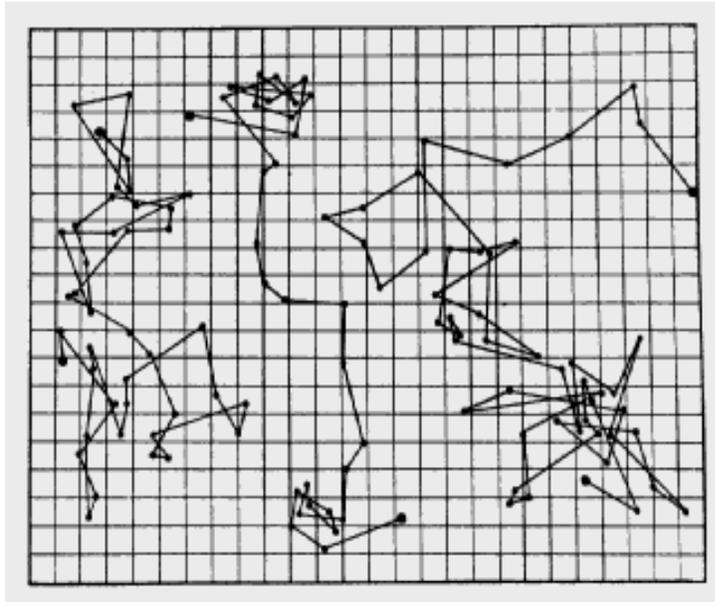}}
\caption{Trajectories of three colloidal particles 
of radius $0.53\mu$m whose positions have been measured experimentally
every 30 seconds. Single points are joined by straight lines
\cite{Perr09}.}
\label{fig:perrin}
\end{figure}

We start this final section with results from a famous
experiment. Fig.~\ref{fig:perrin} shows the trajectories of three
colloidal particles immersed in a fluid. Their motion looks highly
irregular thus reminding us of the trajectory of the drunken sailor's
problem displayed in Fig.~\ref{fig:rwdif}. As we discussed in Chapter
\ref{chap:detdif}, dynamics which can be characterized by a 
normal diffusion coefficient is called Brownian motion. It was
Einstein's achievement to understand the diffusion of molecules in a
fluid in terms of such microscopic dynamics
\cite{Ei05}. His theory actually motivated Perrin to conduct his
experiment by which Einstein's theory was confirmed \cite{Perr09}.

\begin{figure}[t]  
\centerline{\includegraphics[height=12cm,angle=-90]{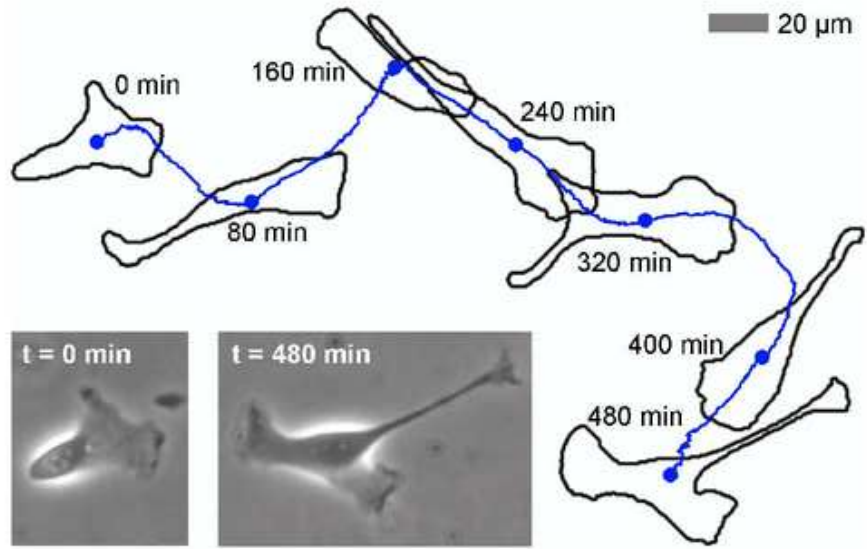}}
\caption{Overlay of a biological cell migrating 
on a substrate. The cell frequently changes its shape and direction
during migration, as is shown by several cell contours extracted
during the migration process. The inset displays phase contrast images
of the cell at the beginning and to the end of its migration process
\cite{DKPS08}.}
\label{fig:cell}
\end{figure}

Fig.~\ref{fig:cell} now shows the trajectory of a very different type
of process. Displayed is the path of a single biological cell crawling
on a substrate \cite{DKPS08}. Nearly all cells in the human body are
mobile at a given time during their life cycle. Embryogenesis,
wound-healing, immune defense and the formation of tumor metastases
are well known phenomena that rely on cell migration. If one compares
the cell trajectory with the one of the Brownian particles depicted in
Fig.~\ref{fig:perrin}, one may find it hard to see a fundamental
difference. On the other hand, according to Einstein's theory a
Brownian particle is {\em passively} driven by collisions from the
surrounding particles, whereas biological cells move {\em actively} by
themselves converting chemical into kinetic energy. This raises the
question whether the dynamics of cell migration can really be
understood in terms of Brownian motion
\cite{DuBr87,SLW91} or whether more advanced concepts of dynamical
modeling have to be applied \cite{HLCC94,URGS01}.

\subsection{Experimental results}

The cell migration experiments that we now discuss have been performed
on two transformed renal epithelial Madin– Darby canine kidney
(MDCK-F) cell strains: wild-type ($NHE^+$) and NHE-deficient ($NHE^-$)
cells.\footnote{$NHE^+$ stands for a molecular sodium hydrogen emitter
that either is present or has been blocked by chemicals, which is
supposed to have an influence on cell migration \cite{DKPS08}.} The
cell diameter is typically 20-50$\mu$m and the mean velocity of the
cells about $1\mu$m/min. The lamellipodial dynamics, which denotes the
fluctuations of the cell body surrounding the cell nucleus, called
cytoskeleton, that drives the cell migration, is of the order of
seconds. Thirteen cells were observed for up to 1000
minutes. Sequences of microscopic phase contrast images were taken and
segmented to obtain the cell boundaries shown in Fig.~\ref{fig:cell};
see Ref.~\cite{DKPS08} for full details of the experiment.

\begin{figure}[t]  
\centerline{\includegraphics[height=12cm,angle=-90]{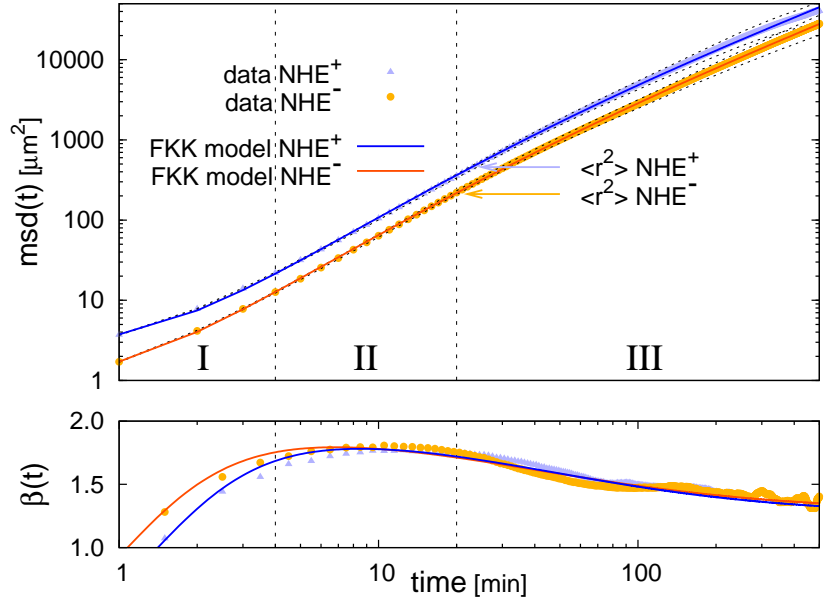}}
\caption{Upper part: Double-logarithmic plot of the mean square 
displacement (msd) as a function of time. Experimental data points for
both cell types are shown by symbols. Different time scales are marked
as phases I, II and III as discussed in the text. The solid lines
represent fits to the msd from the solution of our model, see
Eq.~(\ref{eq:msdn}). All parameter values of the model are given in
\cite{DKPS08}. The dashed lines indicate the uncertainties of the msd
values according to Bayes data analysis. Lower part: Logarithmic
derivative $\beta(t)$ of the msd for both cell types.}
\label{fig:cell_msd}
\end{figure}

As we have learned in Chapter~\ref{chap:detdif}, Brownian motion is
characterized by a mean square displacement (msd) proportional to $t$
in the limit of long time designating normal
diffusion. Fig.~\ref{fig:cell_msd} shows that both types of cells
behave differently: First of all, MDCK-F $NHE^-$ cells move less
efficiently than $NHE^+$ cells resulting in a reduced msd for all
times. As is displayed in the upper part of this figure, the msd of
both cell types exhibits a crossover between three different dynamical
regimes. These phases can be best identified by extracting the
time-dependent exponent $\beta$ of the msd $\sim t^{\beta}$ from the
data, which can be done by using the logarithmic derivative
\be
\beta(t)=\frac{d\ln msd(t)}{d \ln t} \quad .
\ee
The results are shown in the lower part of
Fig.~\ref{fig:cell_msd}. Phase I is characterized by an exponent
$\beta(t)$ roughly below $1.8$. In the subsequent intermediate phase
II, the msd reaches its strongest increase with a maximum exponent
$\beta$. When the cell has approximately moved beyond a square
distance larger than its own mean square radius (indicated by arrows
in the figure), $\beta(t)$ gradually decreases to about $1.4$. Both
cell types therefore do not exhibit normal diffusion, which would be
characterized by $\beta(t)\to 1$ for large times, but move
anomalously, where the exponent $\beta>1$ indicates superdiffusion.

\begin{figure}[t]  
\centerline{\hspace*{-0.5cm}\includegraphics[height=12cm]{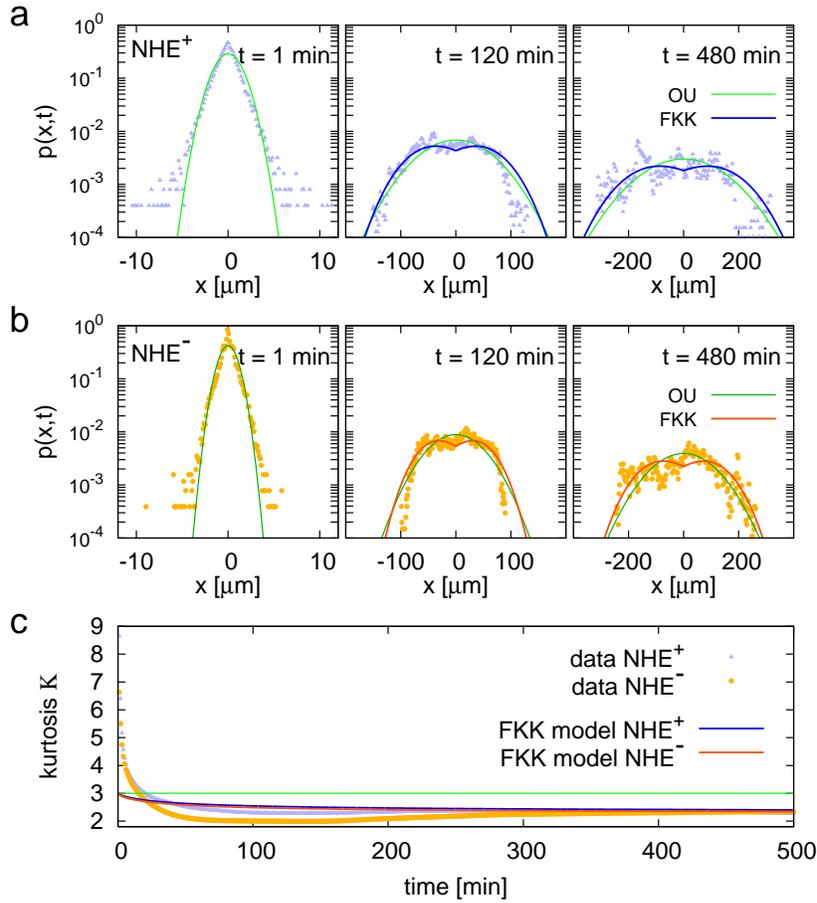}}
\caption{Spatio-temporal probability distributions $P(x,t)$. (a),(b): 
Experimental data for both cell types at different times in
semilogarithmic representation. The dark lines, labeled FKK, show the
solutions of our model Eq.~(\ref{eq:fkk}) with the same parameter set
used for the msd fit. The light lines, labeled OU, depict fits by
Gaussian distributions representing the theory of Brownian motion. For
$t=1$min both $P(x,t)$ show a peaked structure clearly deviating from
a Gaussian form. (c) The kurtosis $\kappa(t)$ of $P(x,t)$, plotted as
a function of time, saturates at a value different from the one of
Brownian motion (line at $\kappa=3$). The other two lines represent
$\kappa(t)$ obtained from our model Eq.~(\ref{eq:fkk})
\cite{DKPS08}.}
\label{fig:cell_pdf}
\end{figure}

We next show the probability that the cells reach a given position $x$
at time $t$, which corresponds to the temporal development of the
spatial probability distribution function
$P(x,t)$. Fig.~\ref{fig:cell_pdf} (a), (b) reveals the existence of
non-Gaussian distributions at different times. The transition from a
peaked distribution at short times to rather broad distributions at
long times suggests again the existence of distinct dynamical
processes acting on different time scales. The shape of these
distributions can be quantified by calculating the kurtosis
\be
\kappa(t):=\frac{<x^4(t)>}{<x^2(t)>^2} \quad ,
\ee
which is displayed as a function of time in Fig.~\ref{fig:cell_pdf}
(c). For both cell types $\kappa(t)$ rapidly decays to a constant
clearly below three in the long time limit. A value of three would be
the result for the spreading Gaussian distributions of the drunken
sailor. These findings are another strong manifestation of the
anomalous nature of cell migration.

\subsection{Theoretical modeling}

We conclude this section with a short discussion of the stochastic
model that we have used to fit the experimental data, as was shown in
the previous two figures. The model is defined by the {\em fractional
Klein-Kramers equation} \cite{BaSi00}
\be
\pard{P}{t}=-\pard{}{x}\left[vP\right]+
\pard{^{1-\alpha}}{t^{1-\alpha}}\gamma\left[\pard{}{v}v+
v_{th}^2\pard{^2}{v^2}\right]P \quad . \label{eq:fkk}
\ee
Here $P=P(x,v,t)$ is the probability distribution depending on time
$t$, position $x$ and velocity $v$ in one dimension,\footnote{No
correlations between $x$ and $y$ direction could be found, hence the
model is only one dimensional \cite{DKPS08}.} $\gamma$ is a damping
term and $v_{th}=k_B T/M$ stands for the thermal velocity of a
particle of mass $M$ at temperature $T$, where $k_B$ is Boltzmann's
constant. The last term in this equation models diffusion in velocity
space, that is, in contrast to the drunken sailors problem here the
velocity is not constant but also randomly distributed according to a
probability density, which is determined by this
equation. Additionally, and again in difference to the simple
diffusion equation that we have encountered in
Section~\ref{sec:diffeq}, this equation features two flux terms both
in velocity and in position space, see the second and the first term
in this equation, respectively. What distinguishes this equation from
an ordinary Klein-Kramers equation, which actually is the most general
model of Brownian motion in position and velocity space \cite{Risk},
is the presence of a {\em Riemann-Liouville fractional derivative} of
order $(1-\alpha)$ in front of the last two terms defined by
\be
\frac{\partial^{\delta} P}{\partial t^{\delta}} := \begin{cases}
\frac{\partial^m P}{\partial t^m} & ,\quad\delta=m \\
\frac{\partial^m}{\partial t^m}\left[\frac{1}{\Gamma (m-\delta)}
\; \int_0^t dt^{\prime }\frac{P(t^{\prime})}{(t-t^{\prime })^{\delta+1-m}}\right] & ,\quad m-1<\delta<m
\end{cases}
\ee
with $m\in\mathbb{N}$. Note that for $\alpha=1$ the ordinary
Klein-Kramers equation is recovered. The analytical solution of this
equation for the msd has been calculated in Ref.~\cite{BaSi00} to
\be
msd(t)=2v_{th}^2t^2E_{\alpha,3}(-\kappa t^{\alpha}) \quad \to\quad
2\frac{D_{\alpha}t^{2-\alpha}}{\Gamma(3-\alpha)}\quad
(t\to\infty) \label{eq:fkkmsd}
\ee
with $D_{\alpha}=v_{th}^2/\gamma$ and the {\em generalized
Mittag-Leffler function}
\be
E_{\alpha,\beta}(z)=\sum_{k=0}^{\infty}\frac{z^k}{\Gamma(\alpha k+\beta)}\:,\:\alpha\:,\:\beta>0\:,\:z\in\mathbb{C} \quad .
\ee
Note that $E_{1,1}(z)=\exp(z)$, hence $E_{\alpha,\beta}(z)$ is a
generalized exponential function. We see that for long times
Eq.~(\ref{eq:fkkmsd}) yields a power law, which reduces to the
Brownian motion result in case of $\alpha=1$. The analytical solution
of Eq.~(\ref{eq:fkk}) for $P(x,v,t)$ is not known, however, for large
friction $\gamma$ this equation boils down to a fractional diffusion
equation for which $P(x,t)$ can be calculated in terms of a Fox
function \cite{SchnWy89}. This solution is what we have used to fit
the data. Our modeling is completed by adding what may be called
``biological noise'' to Eq.~(\ref{eq:fkkmsd}),
\be
msd_{noise}(t):=msd(t)+2\eta^2 \quad . \label{eq:msdn}
\ee
This uncorrelated white noise of variance $\eta^2$ mimicks both
measurement errors and fluctuations of the cell cytoskeleton. The
strength of this noise as extracted from the experimental data is
larger than the measurement error and determines the dynamics at small
time scales, therefore here we see microscopic fluctuations of the
cell body in the experiment. The experimental data in
Figs.~\ref{fig:cell_msd},
\ref{fig:cell_pdf} was then consistently fitted by using the four fit
parameters $v_{th}, \alpha, \gamma$ and $\eta^2$ in Bayesian data
analysis \cite{DKPS08}.

We consider this model as an interesting illustration of the
usefulness of stochastic fractional equations in order to understand
real experimental data displaying anomalous dynamics. However, the
reader may still wonder about the physical and biological
interpretation of the above equation for cell migration. First of all,
it can be argued that the fractional Klein-Kramers equation is
approximately\footnote{We emphasize that only the msd and the decay of
velocity correlations are correctly reproduced by solving such an
equation \cite{Lutz01}, whereas the position distribution functions
corresponding to this equation are mere Gaussians in the long time
limit and thus do not match to the experimental data.} related to the
generalized Langevin equation
\cite{Lutz01}
\be
\dot{v}=-\int_0^tdt'\:\gamma(t-t') v(t')+\sqrt{\zeta}\:\xi(t) \quad .
\ee
This equation can be understood as a stochastic version of Newton's
law: The left hand side holds for the acceleration of a particle,
whereas the total force acting onto it is decomposed into a friction
term and a random force. The latter is modeled by Gaussian white noise
$\xi(t)$ of strength $\sqrt{\zeta}$, and the friction coefficient is
time-dependent obeying a power law, $\gamma(t)\sim t^{-\alpha}$.  The
friction term could thus be written again in form of a fractional
derivative. Note that for $\gamma=const.$ the ordinary Langevin
equation is recovered, which is a standard model of Brownian motion
\cite{Reif,Risk}. This relation suggests that physically the anomalous cell
migration could have its origin, at least partially, in the existence
of a memory-dependent friction coefficient.  The latter, in turn,
might be explained by anomalous rheological properties of the cell
cytoskeleton, which consists of a complex biopolymer gel
\cite{SSGMBK07}.

Secondly, what could be the possible biological significance of the
observed anomalous cell migration? Both the experimental data and the
theoretical modeling suggest that there exists a slow diffusion on
small time scales, whereas the long-time motion is much faster. In
other words, the dynamics displays intermittency qualitatively similar
to the one outlined in Section~\ref{sec:anodif}. Interestingly, there
is an ongoing discussion about optimal search strategies of foraging
animals such as albatrosses, marine predators and fruit flies, see,
e.g., Ref.~\cite{EdWa07} and related literature. Here it has been
argued that L\'evy flights, which define a fundamental class of
anomalous dynamics \cite{KRS08}, are typically superior to Brownian
motion for animals in order to find food sources. However, more
recently it was shown that under certain circumstances intermittent
dynamics is even more efficient than pure L\'evy motion
\cite{BCMV06}. The results on anomalous cell migration presented above
might thus be biologically interpreted within this context.
  
\chapter{Summary}

This review has covered a rather wide range of topics and methods:
{\em Chapter~\ref{prelim}} has introduced to the concept of
deterministic chaos by defining and calculating fundamental quantities
characterizing chaos such as Ljapunov exponents and dynamical
entropies. These quantities were shown to be intimately related to
each other as well as to properties of fractal sets.  {\em
Chapter~\ref{chap:detdif}} started by reminding of simple random walks
on the line, their characterization in terms of diffusive properties,
and the relation to elementary concepts of Brownian motion. These
basic ideas were then discussed in a chaotic setting in form of
deterministic diffusion. A formula was derived exactly expressing
diffusion in terms of the chaos quantities introduced in the previous
chapter.  {\em Chapter~\ref{chap:anodif}} generalized the concept of
normal diffusion leading to anomalous diffusion. A simple
deterministic model generating this type of diffusion was introduced
and analyzed both numerially and by means of stochastic theory. For
the latter purpose, continuous time random walk theory was explained
and applied leading to a generalized, fractional diffusion
equation. As an example of experimental applications of these
concepts, anomalous biological cell migration was
discussed. Experimental results for the mean square displacement and
for the probability distribution in position space matched nicely to
the predictions of a stochastic model in form of a fractional
diffusion equation, defined both in position and velocity space.  The
scope of this review thus spans from very simple, abstract models to
experiments and from basic dynamical systems theory to advanced
methods of stochastic analysis.

Of course, this work poses a wealth of open questions. Here we
restrict ourselves to only one of them which we consider to be of
particular importance: We could understand the origin of normal
diffusion in terms of microscopic deterministic chaos by applying a
combination of methods from statistical physics and dynamical systems
theory; but what's about anomalous diffusion? To our knowledge, yet
there is no analogous theory available, like the escape rate formalism
for normal diffusion, which explains the origin of anomalous diffusion
in terms of weak chaos. This is the reason why here we restricted
ourselves to applying numerical and stochastic methods to an anomalous
deterministic model. We believe that constructing such a microscopic
dynamical systems theory of anomalous deterministic transport poses a
big challenge for future work in this field.\\

{\bf Acknowledgements:}\\ 
The author gratefully acknowledges the long-term collaboration with
J.R.Dorfman leading to the material presented in
Chapter~\ref{chap:detdif}. He also wishes to thank his former Ph.D.\
student N.Korabel for help with some of the figures and for joint work
on Section~\ref{sec:anodif}, which formed part of his Ph.D.\
thesis. A.V.Chechkin significantly contributed to the same section but
in particular introduced the author to the stochastic theory of
anomalous diffusion, for which he is extremely grateful. P.Dieterich
was the main driving force behind the project reviewed in
Section~\ref{sec:anocell}, and the author thanks him for highly
interesting joint work on crawling cells. Finally, he thanks his
former postdoc P.Howard for help with some figures and particularly
for recent first steps towards solving the question posed above.

\newpage

\bibliography{dcad_ref}

\end{document}